# A RECOMMENDER SYSTEM FOR AUTOMATIC PICKING OF SUBSURFACE FORMATION TOPS

A PREPRINT

**Jesse R. Pisel***
Platte River Power Authority, College of Engineering and Applied Science, The University of Colorado at Boulder

**Joshua A. Dierker**
College of Natural Sciences, The University of Texas at Austin

**Sanya Srivastava**
College of Natural Sciences, The University of Texas at Austin

**Samira B. Ravilisetty**
College of Natural Sciences, The University of Texas at Austin

**Michael J. Pyrcz**
Cockrell School of Engineering, Jackson School of Geosciences, The University of Texas at Austin

**Corresponding Author: jesse.pisel@colorado.edu*

February 18, 2022

**ABSTRACT**

Geoscience domain experts traditionally correlate formation tops in the subsurface using geophysical well logs (known as well-log correlation) by-hand. Based on individual well log interpretation and well-to-well comparisons, these correlations are done in the context of depositional models within a stratigraphic framework. However, in the last decade many researchers have focused on automatic well-log correlation using a variety of dynamic time warping algorithms that identify similar parts of pairs of wells, unsupervised machine learning methods that segment individual well-logs and supervised machine learning methods that assign categorical labels based on known tops in many other wells. These methods require a standardized suite of digital well logs (i.e. gamma ray logs for every well) along with the depth to the top of the formations, which might not be available in many cases. Herein, we propose a method that does not use geophysical well logs for correlation, but rather uses already picked tops in multiple wells to recommend the depth to the remaining unpicked tops in the wells. This recommender system calculates the depth to all formation tops in all the wells for two different datasets in two different basins. The Teapot Dome dataset is composed of lithostratigraphic formation tops, and the Mannville Group dataset is composed of sequence-stratigraphic (representing multiple lithologic groups within a stratigraphic unit) formation tops. For the demonstration, mean absolute error and root mean squared error of four-fold cross-validation compares the recommender system predictions to the ground truth human interpretations. The recommender system is competitive and often outperforms state of the art spline interpolation methods. Lastly, increasing the size of the training dataset decreases the prediction error, and that variance in error decreases with increasing formation tops picked in each formation and well for the lithostratigraphic top picks.

***Keywords***: matrix-factorization, formation tops, subsurface, machine learning, lithostratigraphy, sequence stratigraphy

## 1. Introduction

Subsurface datasets usually contain wireline logs, which are used to correlate lithostratigraphic formations, marker beds, facies, and chronostratigraphic units throughout fields (Ainsworth et al., 1999; Wheeler, 2015; Hall, 2016; Gosses, 2020, Pisel and Pyrcz, 2021). Geologists traditionally correlate formations through the subsurface by manual "picking" of tops and bases of each formation, based on regional context and experience with the common spatial patterns associated with depositional settings. This method is useful for integrating expert knowledge with subsurface datasets. However, the subjective nature of interpreting different formation tops means that there is inherent uncertainty in where to pick the correct top, even when chosen by the same interpreter (Doveton, 1994; Lallier et al., 2016; Edwards et al., 2018). Add multiple interpreters to the correlation process, and the lack of repeatability increases significantly. In addition, this manual "picking" method for well-log correlation requires significant human and machine resources (Wu et al., 2018). Choosing the correct depth of tops has volumetric implications for subsurface modeling and development decision making (e.g. carbon sequestration reservoir potential) (Pyrcz and Deutsch, 2014).

In the past decade, automatic well-log correlation has focused on lithostratigraphic correlations via dynamic warping algorithms, supervised machine learning, and neural networks (Wheeler, 2015; Lallier et al., 2016; Edwards et al., 2018; Wu et al., 2018; Brazell et al., 2019; Godwin et al., 2019; Gosses, 2020). The common theme between these methods is warping and pattern matching certain well logs (gamma ray, resistivity, sonic) over multiple scales around the interpreted well tops. However, the methods all require standardized log suites to correlate across the field to basin scale (1 – 100's of km's). These logs are usually digital LAS files and often times pseudodigital image files. Digitizing and normalizing pseudodigital images of logs to digital LAS files is expensive and time consuming, especially for simple tasks such as picking tops to estimate the depth to formation tops. However, formation top data is typically available for many basins from a variety of sources (Wynne et al., 1994; RMOTC (Rocky Mountain Oilfield Testing Center), 2005; Finn, 2019; NAM, 2020). Formation top data is human generated, sparse tabular data that generally contains a unique well identifier, a formation name or unique identifier, and a measured depth (MD) to the top of the formation from a surface datum. Formation top datasets contain picks for each formation in at least one well and typically have picks in hundreds of wells.

Herein we present a new application of collaborative filtering recommendation systems that recommend (calculates) the depth of formation tops without needing any geophysical well logs. This application uses matrix factorization, which can be thought of as a type of dimensionality reduction that moves the dataset into a latent space, where similar points are closer to one another, and minimizes the loss of the regularized predictions in that latent space. This method only needs the subsea depth to the surface, and the unique names of each well. Recommender systems analyze the relationship between two features and attempt to make accurate predictions based on the data collected from these interactions. Many technology companies utilize recommender systems to make recommendations of content to users.

## 2. Data

Teapot Dome is a west-vergent Laramide age basement-cored asymmetric anticline located on the southwestern margin of the Powder River Basin in central Wyoming (Fig. 1 ) (Friedmann and Stamp, 2006). The dome consists of Cambrian through Upper Cretaceous siliciclastic and carbonate sedimentary rocks (Fig. 2)(Friedmann and Stamp, 2006). The dome was first drilled between 1909  and 1915 (Curry Jr, 1977) and has been through primary, secondary, and tertiary production. Additionally, it was owned and operated by the United States Department of Energy (DOE) through 2015 before being sold to private interests (Doll et al., 1995; Friedmann and Stamp, 2006).

Before the sale of Teapot Dome, the DOE along with the Rocky Mountain Oilfield Testing Center (RMOTC) released a public dataset of well logs, depths to lithologic formation tops and well-log correlation marker units, core photographs, 3-D seismic, and production data to encourage scientific discovery and training. This study uses the depths to lithologic formation and member tops and informal marker units (shortened to tops) to evaluate matrix factorization to predict depths for each top in the dataset. There are 1,031 wells in the dataset, and 56 total tops (Figs. 1 and 2). All wells have at least one top, with a mean and standard deviation of six tops per well, and a median of four tops per well. Seventy-five percent of the wells have between three and eight tops. The maximum number of tops is 35 tops in one well. If every well contained every top there would be 67,015 picks (1,031 wells x 65 tops). However there are only 7,286 picks in the dataset meaning that only 10.9 percent of the tops are picked throughout the wells.

The Alberta oil sands region is located in northcentral Alberta on the eastern side of the western Canada sedimentary basin (Fig. 3). The oil sands region consists of Paleozoic through Lower Cretaceous carbonate and siliciclastic sedimentary rocks (Wynne et al., 1994). The area was first drilled in 1906 for conventional oil deposits and bitumen sands were mapped in 1913 (Humphries, 2008). Development of the oil sands bitumen resource has continued in some fashion since (Wynne et al., 1994; Humphries, 2008). In the mid-80’s through mid-90’s the Alberta Geological Survey undertook an effort to map and analyze wells in the area to determine if gas production influences heavy oil production in the region, while storing the data in a relational database (Wynne et al., 1994). Wynne et al. (1994) use a sequence stratigraphic framework for picking transgressive and erosional surfaces in the Lower Cretaceous Mannville

Group in many of the wells (Wynne et al., 1994). The dataset consists of the measured depth to erosional and transgressive surfaces correlated throughout the study area (Wynne et al., 1994).

This sequence stratigraphic dataset is publically available from the Alberta Geological Society and used in a machine learning model for predicting the depth to picks (Wynne et al., 1994; Gosses, 2020). There are 2,122 wells in the dataset and 14 total tops. All wells have at least one top, with a mean, median, and standard deviation of nine, ten, and two tops per well respectively. Seventy-five percent of the wells have between nine and ten tops. The e10 top is the only top without any picks throughout the data subset. The maximum number of tops is 12 tops in one well. If every well contained every top there would be 29,708 picks (2,122 wells x 14 tops). However, there are only 20,328 picks in the dataset meaning that 68.4 percent of the tops are picked throughout the wells. To pick the remaining 89.1 percent of the tops for the Teapot Dome dataset and the remaining 31.6 percent of the tops for the Mannville Group, we propose building a recommender system using matrix factorization with collaborative filtering and alternating least squares.

## 3. Machine Learning

Matrix factorization uses collaborative filtering and alternating least squares (ALS) (Koren et al., 2009). The algorithm works as follows: first, it initializes the user latent vectors and item latent vectors with random values. Next, it then holds the item latent vectors constant, and minimizes the loss for the user latent vectors. After minimizing for the user latent vectors, the algorithm updates the user latent vectors and proceeds to hold them constant while minimizing for the item latent vectors. It then updates the item latent vectors and alternates between the two sets of latent vectors until reaching a minima or specified number of iterations (Koren et al., 2009)(Fig. 4). This is either run for train-test splits or used in k-fold cross-validation for a dataset.

K-fold cross-validation is where the dataset is iteratively split into test and train subsets and uses to train and test the model (Arlot et al., 2010). The dataset is split into four equal subsets (25 percent each). Then a model is trained on three of the subsets (75 percent of the total dataset) and the predictions validated against the known values in the withheld portion (the model has not seen). This is the first of the four folds. Next, three subsets are selected for training, this time including the previously withheld subset, and two of the subsets used in fold 1. The model is then trained from the beginning on these three subsets, and validated on the remaining subset that was not used in training for this fold. This is the second of the four folds. The remaining two folds proceed in a similar fashion on normalized datasets.

## 4. Methods

The first assumption in data normalization for this study is that the wells are vertical with negligible deviation. In addition, the calculations assume that any changes in apparent stratigraphic thickness due to structural dip are fully captured in the dataset. One last key assumption is that the human-picked tops are correct, and have been visually inspected for consistency. With these assumptions laid out, one can assume that the measured depth (*MD*) is approximately equal to the true vertical depth (*TVD*). With this assumption, *TVD* of each top is calculated using the MD and the Kelly bushing (*KB*) or drill floor (*DF*) height:

$$TVD = MD - KB. \quad (1)$$

This standardization ensures that the true vertical depth of each formation top is relative to the ground surface elevation. It also serves as quality control to identify any incorrectly entered information in the dataset (human errors). Next, from *TVD* the true vertical depth relative to sea level (*TVDSS*) is calculated by subtracting elevation (*EL*) of the ground surface at the well from TVD:

$$TVDSS = TVD - EL. \quad (2)$$

This serves to ensure that the dataset has stationarity, because the ground surface elevation varies across the Teapot Dome field due to Neogene erosion. Lastly, all *TVDSS* values are normalized by subtracting the minimum *TVDSS* from each top to ensure that all values are positive. This makes the latent factors easier to inspect, and the results are converted back to TVDSS. After normalizing the tops, the dataset is then used with matrix factorization and ALS.

The ALS algorithm performs well with large sparse datasets, which is the case with both the Teapot Dome and Mannville Group datasets. ALS matrix factorization for these datasets is simply reconfigured so that the wells (users) and tops (items) are the two factors. So for example, to calculate the recommended depth of unknown tops ($\hat{r}_{ui}$) for well *u* and formation top *i* is as follows:

$$\hat{r}_{ui} = q_i^T p_u \quad (3)$$

where $q_i^T$ is the top latent vector, and $p_u$ is the well latent vector (Fig. 5). ALS minimizes the loss function from Koren et al. (2009) of the form:

$$L = \sum_{(u,i)\in k} (r_{ui} - q_i^T \cdot p_u)^2 + \lambda( \parallel q_i \parallel^2 + \parallel p_u \parallel^2) \quad (4)$$

The loss function, *L*, is an error metric that compares the squared difference between the actual top depth ($r_{ui}$) and the recommended top depth, and adds an *L2* normalization with hyperparameter ($\lambda$) to prevent overfitting of the well and top latent vectors (Koren et al., 2009). Mathematically the right hand side of the loss function can be thought of as two separate parts. The first part is the squared difference between the predicted value and the actual value for known cases. This is the measure of how close the predicted top depths are to the known top depths. A value of zero means there is perfect agreement between the predicted depth and the actual depth.

The second part of the right hand side of the equation deals with overfitting. Overfitting can be thought of a model that fits noise in a dataset as if that noise were part of the underlying structure of the data. For this problem, overfitting means that one or two formation tops would have low error, while the remaining tops would have high error values. Regularization helps the model generalize the depths to each formation top. The L2 normalization is composed of a user-defined hyperparameter multiplied with the squared magnitude of the top and well latent vectors. Meaning that as the error of the prediction becomes smaller; the loss approaches the value of λ multiplied by the squared magnitudes of the latent vectors. Which in turn means the solution is more generalized and is not over fit to the data points with known values. The minimization of the loss function is important because the smaller the loss, the smaller the error in the recommendation.

Choosing an optimal number of factors for the latent vectors is directly related to complexity of the original dataset and the level of granularity in the recommendations. Lower number of factors will recommend the depths of the most common tops, but will not fit accurately to individual wells. However, too many factors leads to overfitting, which is counteracted by the regularization term above (Jannach et al., 2013). Tuning this hyperparameter along with two other hyperparameters is important to minimize the prediction error.

Matrix factorization with ALS uses the following hyperparameters: the number of latent factors, the number of iterations to run the algorithm, and the regularization parameter. The hyperparameters are optimized via an exhaustive grid search on the following: one to 10 latent factors, 10 to 440 iterations in steps of 10, and 5 different regularization parameters ($\lambda$) from 0.001 to 10. To validate each of these 2,200 combinations, four-fold cross-validation is used to measure the accuracy of the predictions against human interpreted top picks. The picks included in each fold are selected at random, with each pick included in only one validation fold, with the exception of tops that only have one pick being included in all four folds (otherwise the matrix factorization will not make predictions for these single picks). A spatially blocked four-fold cross validation (Roberts et al., 2017; Uieda, 2018) is also used to measure the influence of spatial patterns in prediction error. Four folds are used because it is computationally efficient, and yields slightly over 5,000 picks per fold for training. This is analogous to having 200 unique wells and 50 tops picked throughout half of the wells, which is the size of a relatively small onshore field.

In addition to cross-validation, the dataset is split into training and testing subsets to evaluate the accuracy of the model with different sizes of training data. For this method, the data is split into train and test subsets that range from 1 percent to 99 percent of the full dataset. At each iteration the model trains on the training subset; accuracy is then measured on the test subset. After this, the model returns to the initial random starting state, and trains for 99 more iterations. At the end of 100 iterations, the ratio of test to train is increased by 10 percent. This process continues until the ratio of train to test is 99 percent to 1 percent respectively.

To evaluate each of the 354,000 training runs (2,200 combinations by four folds random plus 2,200 combination by four folds spatially blocked times 2 datasets plus 2,000 iterations of increasing the train-test split ratio) the method will use the mean absolute error (MAE) and root mean squared error (RMSE) metrics. These metrics are both easily interpretable numbers, i.e. the predicted top depth is 3 m away from the actual top depth, and provide insight into our predictions.

Lastly, the depth predictions (recommendations) are compared to current state of the art methods using an automatic Green's function spline interpolation in Verde (Uieda, 2018). The spline interpolation undergoes the same four-fold cross-validation process both random as well as spatially blocked. The MAE and RMSE are compared for both the Teapot Dome and Mannville Group datasets for every formation top pick. In addition, the difference between the two methods error metrics is used to measure the performance of the two methods relative to one another. For this error difference, one simply subtracts the recommender system error from the spline interpolation error. Positive values document the recommender system has lower error than the spline interpolation, while negative values document the spline interpolation has lower error than the recommender system.

## 5. Results

The optimal hyperparameters for the matrix factorization model trained on the Teapot Dome dataset are as follows: two latent factors, 290 iterations, and a regularization parameter ($\lambda$) of 0.1. The MAE for the optimized model ranges from 5.4 to 6.9 meters (m) with a MAE of 6.8 m averaged across the four folds. The RMSE ranges from 7.0 to 9.7 m with an average of 8.9 m across the four folds (Table 1). The MAE for the optimized model with spatially blocked folds ranges from 21.7 to 39.7 m with an average of 28.1 m. The RMSE ranges from 23.7 to 30.0 m with an average of 27.0 m across the four folds (Table 1).

The mean absolute error for each top in each fold is summarized in Appendix 1. The minimum MAE for all folds and all tops is 0 m for the C3 Sand member of the Tensleep Sandstone in fold 2 (n train = 4, n test = 1), which also has the lowest RMSE of 0.0 m as well. The maximum MAE across all folds and all tops is 50.4 m for the upper member of the Sundance Formation in fold 2 (n train = 4, n test = 3), which also has the highest RMSE of 72.0 m (Appendix 1). Averaged across the four folds, the C4 Sand member of the Tensleep Sandstone has the lowest MAE and RMSE errors both of 0.8 m (n picks = 5). The maximum MAE and RMSE averaged across the four folds is the D Dolomite member of the Tensleep Sandstone with errors of 46.5 and 46.4 m respectively (n picks = 2).

For non-spatial four-fold cross-validation, the recommender system has lower MAE and RMSE errors than the spline interpolation for all formation tops except the F1WC base and top of the Carlile Shale (Appendix 1). On average, the difference between the spline interpolation and the recommender system MAE is 38.2 m and RMSE is 47.7 m, which favors the recommender system.

For spatially-blocked cross-validation, the minimum MAE for all folds and all tops is 1.5 m for the top of the Sussex Sandstone of the Cody Shale in folds 1 and 3 (n train = 200, n test = 64), while the Muddy Sandstone in fold 2 has the lowest RMSE of 2.3 m. The maximum MAE across all folds and all tops is 194.1 m for the C3 Dolomite member of the Tensleep Sandstone in fold 4 (n train = 3, n test = 1), which also has the highest RMSE of 194.1 m (Appendix 1). Averaged across the four spatially-blocked folds, the Sussex Sandstone of the Cody Shale has the lowest MAE and RMSE errors of 1.8 and 3.1m (n picks = 264). The maximum MAE and RMSE averaged across the four folds is the C4 Sand member of the Tensleep Sandstone with errors of 113.6 and 114.1 m respectively (n picks = 5).

For the spatially blocked four-fold cross-validation, the recommender system has lower MAE errors for 60 percent of the tops, and lower RMSE errors for 70 percent of the tops than the spline interpolation (Appendix 1). On average, the difference between the spline interpolation and the recommender system MAE is -7.6 m and RMSE is -0.7 m, which favors the spline interpolation.

On a well-by-well basis, the average MAE per well is 3.6 m and the RMSE is 4.1 m. There are 257 wells with total error less than 0.3 m. The maximum MAE is 127.8 m in the NPR #3 13-StX-23-H. Spatially, the MAE appears randomly distributed throughout the field (Fig. 6). In addition to four-fold cross validation, the size of the training and validation datasets are varied. The MAE decreases as the size of the training dataset increases. A training dataset of 1 percent of the total dataset yields a MAE of ~750 m, while a training dataset with 99 percent of the total dataset yields a MAE of ~5 m (Fig. 7).

Optimal hyperparameters for the matrix factorization model trained on the Mannville Group dataset are as follows: three latent factors, 100 iterations, and a regularization parameter (λ) of 0.1. The MAE for the optimized model ranges from 9.5 to 11.8 m with a MAE of 10.6 m averaged across the four folds. The RMSE ranges from 24.5 to 35.3 m with an average of 27.4 m across the four folds (Table 2). The MAE for the optimized model with spatially blocked folds ranges from 11.4 to 13.2 m with an average of 11.7 m. The RMSE ranges from 20.8 to 40.5 m with an average of 27.6 m across the four folds (Table 2).

In the Mannville Group predictions, the minimum MAE for all folds and all tops is 2.6 m for Regression 2 in the Wabiskaw member of the Clearwater Formation in fold 1 (n train = 1,382, n test = 449), while the top of the Clearwater Formation/Wabiskaw member has the lowest RMSE of 4.2 m. The maximum MAE across all folds and all tops is 43.6 m for the top of the Mannville Group in fold 4 (n train = 1,195, n test = 396), which also has the highest RMSE of 170.7 m (Appendix 1). Averaged across the four folds, the top of the Clearwater Formation/Wabiskaw member has the lowest MAE and RMSE of 3.7 and 5.3 m (n picks = 1,831). The maximum MAE and RMSE averaged across the four folds is the top of the Mannville Group with MAE and RMSE of 40.7 and 82.7 m respectively (n picks = 1,591).

For non-spatial four-fold cross-validation, the recommender system has lower MAE than the spline interpolation for all tops. The RMSE errors only contain two instances where the spline interpolation error is lower than the recommender system and those are the top of the Mannville Group and top of the Paleozoic (Appendix 1). On average, the difference between the spline interpolation and the recommender system MAE is 16.2 m and RMSE is 72.1 m, which favors the recommender system.

For spatially-blocked cross-validation, the minimum MAE for all folds and all tops is 3.1 m for the top of Regression 2 in the Wabiskaw member of the Clearwater Formation in fold 2 (n train = 1,379, n test = 452), while the top of the Clearwater Formation/Wabiskaw member has the lowest RMSE of 4.7 m in fold 1. The maximum MAE across all folds and all tops is 47.1 m for the top of the Mannville Group in fold 4 (n train = 1204, n test = 387), which also has the highest RMSE of 132.6 m (Appendix 1). Averaged across the four spatially-blocked folds, the top of the Clearwater Formation/Wabiskaw member has the lowest MAE and RMSE errors of 3.8 and 6.2 m (n picks = 425). The maximum MAE and RMSE averaged across the four folds is the top of the Mannville Group with errors of 72.9 and 73.0 m respectively (n picks = 1591).

For the spatially blocked four-fold cross-validation, the recommender system has lower MAE errors for 71 percent of the tops, and lower RMSE errors for 92 percent of the tops than the spline interpolation (Appendix 1). On average, the difference between the spline interpolation and the recommender system MAE is 14.4 m and RMSE is 142.6 m, which favors the recommender system.

On a well-by-well basis, the average MAE per well is 11.6 m and the RMSE is 14.3 m. There are 126 wells with total error less than 0.3 m. The maximum MAE and RMSE is 1,131 m and 1,885 m in the 00/06-30-071-09W4/0 well. Spatially, the MAE appears randomly distributed throughout the field (Fig. 8).

In addition to four-fold cross validation, the size of the training and validation datasets is varied. The MAE decreases as the size of the training dataset increases. A training dataset of 1 percent of the total dataset yields a MAE of ~100 m, while a training dataset with 99 percent of the total dataset yields a MAE of ~10 m (Fig. 7).

In comparison to the XGBoost model of Gosses (2020) for predicting the top of the McMurray Formation in the Mannville Group dataset, the recommender system averages a MAE of 6.4 m and a RMSE of 23.8 m across the four non-spatial cross validation folds. Across the spatially blocked folds, the model MAE and RMSE is 6.4 m and 23.9 m respectively.

## 6. Discussion

The optimal hyperparameters for both models are a low number of latent factors (2 and 3) which means that the problem of recommending top depths is easily solved in a two to three-dimensional latent space. Low dimension latent space solutions are interpreted to mean that the problem is not too challenging to learn. Notably the lithostratigraphic picks were optimal with two latent factors, and the sequence stratigraphic picks were optimal with three latent factors. Intuitively this makes sense, in that sequence stratigraphic tops are more challenging for humans to pick than lithostratigraphic tops and requires a higher dimension latent space to learn the depths.

The number of iterations for both models is less than 300, and this simply relates to how quickly the models converge to a global minima. The lithostratigraphic model takes more iterations to converge to the global minima, while the sequence stratigraphic model converges after 100 iterations. This is likely due to the difference in data sparsity between the two datasets. The Teapot Dome dataset only has 10 percent of the total tops picked throughout the dataset, while the Mannville Group dataset has 68 percent of the total tops picked. The sparser the dataset, the longer it takes the model to find the global minima with ALS as it has to predict more missing values.

For the *L2* regularization, hyperparameter ($\lambda$) a value of 0.1 for both datasets is reasonable. As a quick recap, as the value of $\lambda$ increases so does the amount of bias in the predictions. With high values of $\lambda$, the models begin to under fit to the data, while low values of $\lambda$ negate the regularization term and results in overfitting. Values of zero remove the regularization term altogether. With this in mind, a value of 0.1 is interpreted as similar variance in both datasets that is counteracted by the regularization. This variance could be attributed to the vertical well assumption or by differences in human picks across both datasets. More work is needed to evaluate the regularization for these datasets and the impact it has on the RMSE and MAE on a formation top basis and on a well-by-well basis.

For the Teapot Dome predictions, the tops with the largest RMSE and MAE are all tops that have less than 10 tops picked in the original dataset (Appendix 1). Plotting the error values against the number of human picks for each top, there is a decrease in the error variance as the number of tops increases (Fig. 9). This makes intuitive sense, as more information is introduced into the recommender system its predictions become more accurate. However, the Mannville Group predictions does not have a similar trend. The number of top picks does not appear to decrease the MAE or RMSE and their variances (Fig. 9). If the Mannville Group dataset were sparser the decrease in error variance might be apparent, but more work is needed to verify if the decrease in error variance is directly related to data sparsity or underlying difference between lithostratigraphic picks and sequence stratigraphic picks. Analyzing the error from the different

Mannville Group tops, there is no difference between the error for transgressive and regressive top picks (Fig. 9). This means that interpreters consistently picked these surfaces. For example, if there was more error in the transgressive top picks, one could attribute this to the human interpreters not consistently picking the same geophysical response as it changes from well to well. Because of the similarity in error between the regressive and transgressive surfaces, the human interpretations are not likely a major source of uncertainty in this dataset.

Comparing the non-spatial four-fold cross-validation results with state of the art spline interpolation for both the Teapot Dome and Mannville Group predictions, the recommender system overwhelmingly outperforms the interpolation with lower MAE and RMSE values. This is especially true for tops that have fewer than 10 picks in the Teapot Dome dataset. The spline interpolation uses location and depth to a top to interpolate to a regular grid while the recommender system uses all of the other top information to make predictions. This means that the interpolation is using less than 10 data points to make predictions. The recommender system however leverages all tops in all wells, so while the individual formation may only have 10 picks, the system is using up to 7,286 data points to make predictions. Again, more data means better predictions and lower MAE and RMSE values. In the Mannville Group the recommender system outperforms the interpolation as well, which suggests that the spatially discontinuous nature of sequence stratigraphic picks might be better suited to this method than a minimum curvature spline that is interpolating through areas where surfaces converge, or are removed by erosion.

This hypothesis is tested by the spatially blocked four-fold cross-validation. When the folds are spatial blocks, the spline interpolation on average outperforms the recommender system for the Teapot Dome prediction. Ironically, the tops with low number of picks that were an asset for the recommender system during random four-fold cross-validation have the highest error. This suggests that when spatial blocks are missing from a lithostratigraphic top dataset the recommender system simply predicts the average depth of the tops. With the Mannville Group predictions, the recommender system outperforms the spline interpolation on average, but the MAE for the spline is lower for tops that have less than 300 picks. However, the recommender system has a lower RMSE for all predictions except for Transgression 6. For the sequence stratigraphic dataset, the spatial blocks appear to be less important for the recommender system to make predictions with low MAE and RMSE. One easy solution to deal with the lack of spatial information in the recommender system is to use factorization machines, which are a higher-order form of matrix factorization that uses additional features such as spatial location, human interpreter and other well specific information.

On a well-by-well basis, there are no systematic trends in the MAE or RMSE. One possible explanation was correlation between features in the dataset (e.g. number of picks, operator, status, class, total depth, datum, datum elevation, ground elevation, completion and spud dates) and the error. There is no correlation between any features and the error on a by-well basis. The features listed do not influence the picked depth for the tops in the dataset. Next, there is no correlation between geologic age of tops and error, and the predictions have no correlation with which tops were picked and the error. For example, in the Teapot Dome predictions the 10 wells with the highest RMSE and MAE and the 10 wells with the lowest RMSE and MAE all had less than 20 picks, and were exclusively in the Cretaceous. The only item to note on a well-by-well basis is that the variance of the error increases in wells that have fewer picks, which makes sense, as there is less information in those wells for the recommender system to use for making predictions (Fig. 6). This is reinforced by results from the Mannville Group predictions where error and variance is inversely related to the number of picks per well (Fig. 8).

In comparison to the published model of Gosses (2020), the recommender system does not perform as well as the XGBoost model. The XGBoost model has a RMSE of 6.66 m compared to the 23.9 m RMSE of the recommender system for predicting the top of the McMurray Formation. However, the tops dataset is less than 2 megabytes in size compared to the 390-megabyte dataset required to train the XGBoost model. Additionally, as discussed in the background section, the tops dataset requires minimal data preparation and munging compared to the well-log dataset. Additionally, the recommender system model is learning the depth of each formation top in each well relative to one another and not the well-log pattern. This suggests that it should generalize well for picking both lithostratigraphic and sequence stratigraphic tops. Overall, the recommender system does relatively well on these datasets. The small size of the tops datasets, training speed, and minimal data preprocessing means this method is both compute and time efficient. It would best be used in conjunction with other machine-learning methods to select an initial depth for each top before refining the depth with the well log.

More work is needed to apply this method to a wider variety of datasets, and more quantitative work is needed to determine the specific sources of error in the predictions, such as interpreter bias and spatial dependence of the error. Additionally, the main assumption of this article is that the wells are all vertical. If the TVDSS for each human picked top was corrected using the wellbore deviations, this could decrease the error in both datasets. One of the interesting benefits of this method is because it is learning the subsea depths of picks instead of the geophysical response of the rocks; it will actually predict the structural depth of tops that have been removed by erosion. To the matrix factorization model, the tops that are missing by erosion are simply missing tops to predict, and it treats them just the same as any other unpicked tops in the subsurface. In the case of Teapot Dome, it predicts the upper Cretaceous tops removed by erosion. This means that the method could be useful in constraining erosion rates and 3D geometries of structural features.

## 7. Conclusion

In conclusion, matrix factorization is competitive with state of the art interpolation methods for predicting TVDSS for both lithologic and sequence stratigraphic surfaces. Spatial and non-spatial four-fold cross-validation documents that this method outperforms spline interpolation methods when tops are missing at random. While the method does not perform as well as other machine learning methods on the sequence stratigraphic dataset, it is a lightweight first pass solution to automatically picking formation tops in the subsurface. Future work on this topic should include evaluating bias in the underlying datasets, spatial dependence of the error metrics, adding higher dimensions into the matrix factorization such as spatial locations, and the effect of repeated or missing tops in faulted and folded areas.

## Acknowledgements

The authors would like to thank ConocoPhillips for providing funding for this research through the Freshman Research Initiative Energy Analytics stream, which is part of the College of Natural Sciences at the University of Texas at Austin. The authors would like to thank Peter Burgess and Justin Gosses for reviewing different versions of this manuscript and providing valuable insight that increased the quality of this manuscript. The authors would also like to thank four additional anonymous reviewers for reading this manuscript as well.

### Data and code source

The original Teapot Dome lithostratigraphic picks are available at http://s3.amazonaws.com/open.source.geoscience/open_data/teapot/rmotc.tar. The original Mannville Group sequence stratigraphic picks are available at https://ags.aer.ca/publications/SPE_006.html. Code used to munge the datasets, complete the analysis, and plot the results, along with the munged datasets are available at https://zenodo.org/record/3990398. Code development is active at https://github.com/jessepisel/matrixfactorization.

## FIGURE CAPTIONS

**Figure 1.** Location map of the Wyoming Laramide age sedimentary basins, the Teapot Dome field, and inset map of well locations used in this study.

**Figure 2.** Stratigraphic ages, position, formation, member, informal names, and number of picks per top for every top in the Teapot Dome dataset.

**Figure 3.** Above: Stratigraphic ages, position, formation, member, and informal names, along with the number of picks for each top in the Mannville Group dataset. Below: The inset map is the location of the wells used in this portion of the study in west-central Alberta, Canada.

**Figure 4.** Flowchart for implementing matrix factorization with alternating least squares.

**Figure 5.** Schematic representation of how matrix factorization works. The formation top latent vectors are multiplied by the well latent vectors and error is minimized at each iteration.

**Figure 6.** Error maps for the Teapot Dome dataset for both the MAE and RMSE. There is no spatial patterns in the error per well (left).Variance in error decreases as the number of picks per well increases (right).

**Figure 7.** Plot of the mean absolute error (MAE) in meters for the Teapot Dome and Mannville Group datasets, against the fraction of the dataset used in training.

**Figure 8.** Error maps for the Mannville Group dataset for both the MAE and RMSE. There appears to be small areas of larger error in the southwest and northwest areas of the study area (left). The variance in error does not decrease as the number of picks per well increases (right). However, the average error does decrease as the number of picks per well increases.

**Figure 9.** Scatter plot of error and the number of picks for each formation. Each individual point is representative of each formation in the datasets. In the Teapot Dome plot, formation errors are colored by geologic age in Fig. 2. In the Mannville Group plot, regressive surfaces are orange, and transgressive surfaces are green as depicted in Fig. 3.

**Table 1.** Error for the cross-validation folds on the Teapot Dome dataset.

**Table 2.** Error for the cross-validation folds on the Mannville Group dataset.

**Appendix 1.** Analysis results for the spatial and non-spatial four cross-validation folds. Rows are results for each formation top, columns are repeated within the folds and contain number of training samples, number of holdout samples, mean absolute error (m) and root mean squared error (m). Spline interpolation results are also aligned with the individual formation tops. Comparison for random cross validation, and spatial-blocked cross validation.

# Location Map of the Teapot Dome Dataset

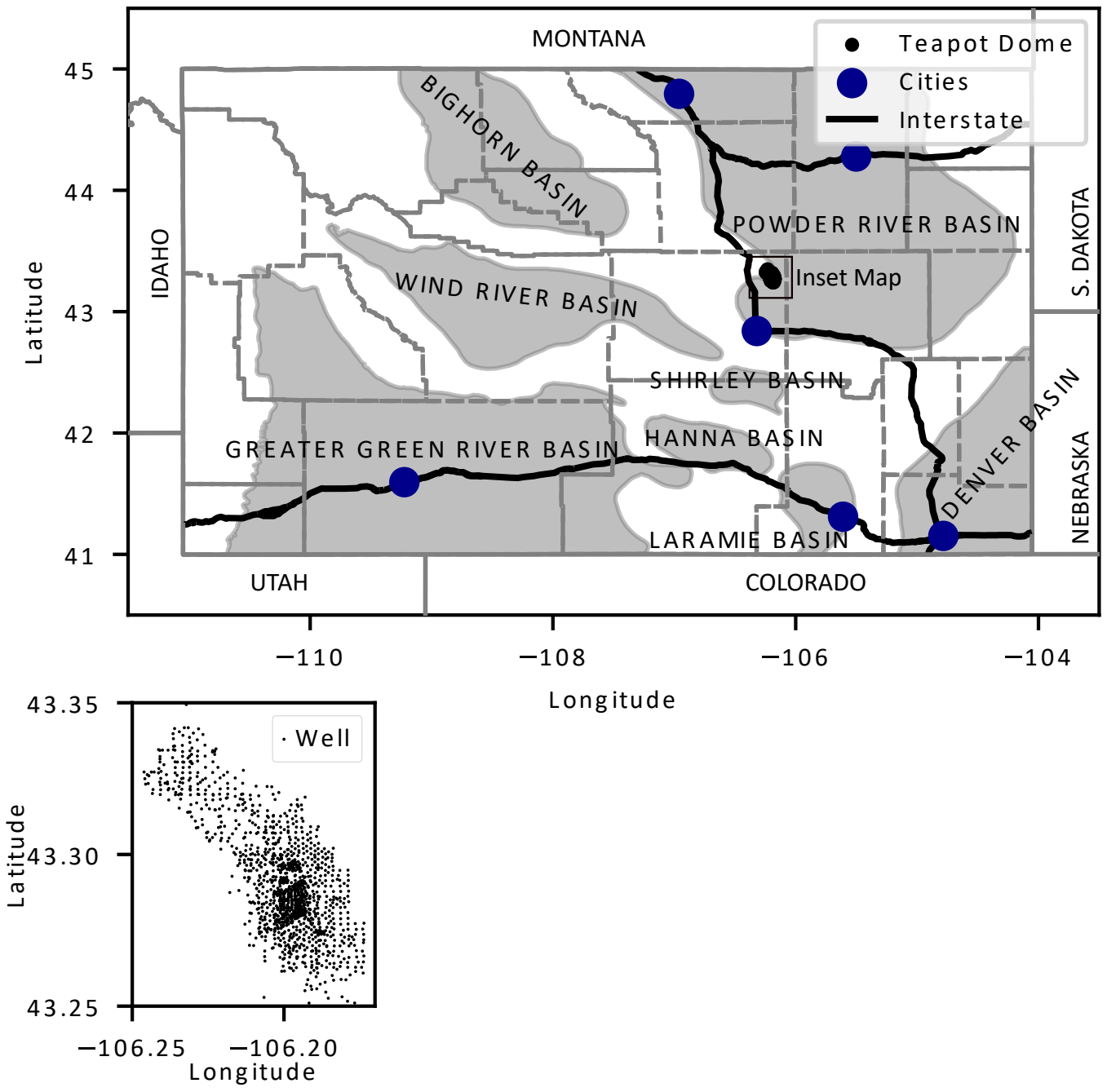

# Teapot Dome Stratigraphic Column

| System | Formation | Color | Code | Top Description | Number of Picks |
|---|---|---|---|---|---|
| Cretaceous – Upper Cretaceous | Cody Shale – Steele Shale Mbr. | | SSXS | Sussex Sandstone | 264 |
| | | | SSXbase | Sussex Sandstone Base | 351 |
| | | | SHNNu | Upper Shannon Sandstone | 916 |
| | | | SHNNl | Lower Shannon Sandstone | 920 |
| | | | SHNNbs | Shannon Sandstone Base | 669 |
| | | | StBR | Steele Shale Brittle Zone | 100 |
| | | | StFT | Steele Shale Fishtooth | 100 |
| | | | StGD | Steele Shale Gray Dust | 90 |
| | | | StAM | Steele Shale Ardmore Bentonite | 108 |
| | Niobrara Mbr. | | NBRRws | Niobrara White Specks | 189 |
| | | | NBRRsg | Niobrara Smokey Gap Member | 188 |
| | Carlile Sh. | | CRLL | Carlile Shale | 268 |
| | Frontier Formation | | F1WC | 1st Frontier | 332 |
| | | | F1WCBench1Base | 1st Frontier 1st Bench Base | 95 |
| | | | F1WCBench2Top | 1st Frontier 2nd Bench | 88 |
| | | | F1WCBench2Base | 1st Frontier 2nd Bench Base | 73 |
| | | | F1WCBench3Top | 1st Frontier 3rd Bench | 74 |
| | | | F1WCbase | 1st Frontier Base | 332 |
| | | | B1 | B1 Marker | 276 |
| | | | F2WC | 2nd Frontier | 324 |
| | | | F2WCbase | 2nd Frontier Base | 289 |
| | | | F3WC | 3rd Frontier | 54 |
| | | | F3WCbase | 3rd Frontier Base | 44 |
| | Mowry Sh. | | MWRY | Mowry Shale | 53 |
| Lower | Muddy Ss. | | MDDY | Muddy Sandstone | 31 |
| | Thermopolis | | THRM | Thermopolis Shale | 29 |
| | Dakota Ss. | | DKOT | Dakota Formation | 46 |
| | Lakota Ss. | | LKOT | Lakota Formation | 31 |
| Jurassic | Morrison Fm. | | MRSN | Morrison Formation | 31 |
| | Sundance Formation | | SNDCu | Upper Sundance Formation | 7 |
| | | | CNSP | Canyon Springs Member | 5 |
| Triassic | Chugwater Formation | | CRMT | Crow Mountain Member | 18 |
| | | | ALCV | Alcova Limestone Member | 15 |
| | | | RDPK | Red Peak Member | 15 |
| Permian | Goose Egg Formation | | ERVY | Ervay Evaporite Member | 2 |
| | | | FRLL | Forelle Limestone Member | 3 |
| | | | GLND | Glendo Shale Member | 1 |
| | | | MNKT | Minnekahta Limestone Member | 36 |
| | | | OPCH | Opeche Shale Member | 34 |
| Pennsylvanian | Tensleep Sandstone | | A Sand | Tensleep A Sand | 33 |
| | | | B Dolo | Tensleep B Dolomite | 35 |
| | | | B Sand | Tensleep B Sand | 35 |
| | | | C1 Dolo | Tensleep C1 Dolomite | 32 |
| | | | C1 Sand | Tensleep C1 Sand | 2 |
| | | | C2 Dolo | Tensleep C2 Dolomite | 2 |
| | | | C2 Sand | Tensleep C2 Sand | 5 |
| | | | C3 Dolo | Tensleep C3 Dolomite | 4 |
| | | | C3 Sand | Tensleep C3 Sand | 5 |
| | | | C4 Dolo | Tensleep C4 Dolomite | 5 |
| | | | C4 Sand | Tensleep C4 Sand | 5 |
| | | | D Dolo | Tensleep D Dolomite | 2 |
| | | | D Sand | Tensleep D Sand | 1 |
| | | | E Dolo | Tensleep E Dolomite | 1 |
| | Amsden Fm. | | AMSD | Amsden Formation | 1 |
| Mississippian | Madison Ls. | | MDSN | Madison Limestone | 1 |
| Pre-Cambrian | Pre-Cambrian | | PC | Pre-Cambrian Basement | 1 |

# Mannville Group Stratigraphic Column and Location Map

| System | Formation | | | Color | Code | Top Description | No. of Picks |
|---|---|---|---|---|---|---|---|
| Cretaceous | Mannville Group | Grand Rapids Formation | | | MG | Mannville Group | 1591 |
| | | | | | t61 | Transgression 8 | 328 |
| | | | | | t51 | Transgression 7 | 362 |
| | | Clearwater Formation | | | t41 | Transgression 6 | 588 |
| | | | | | t31 | Transgression 5 | 1908 |
| | | | Wabiskaw Member | | CW | Clearwater/Wabiskaw | 425 |
| | | | | | t21 | Transgression 4 | 1576 |
| | | | | | e20 | Regression 3 | 0 |
| | | | | | t15 | Transgression 3 | 1783 |
| | | | | | e14 | Regression 2 | 1783 |
| | | | | | t11 | Transgression 2 | 1831 |
| | | | | | t10.5 | Transgression 1 | 1797 |
| | | | | | e10 | Regression 1 | 1783 |
| | | McMurray Formation | | | MMF | McMurray Formation | 1880 |
| Undifferentiated Paleozoic | | | | | PZ | Paleozoic rocks | 2099 |

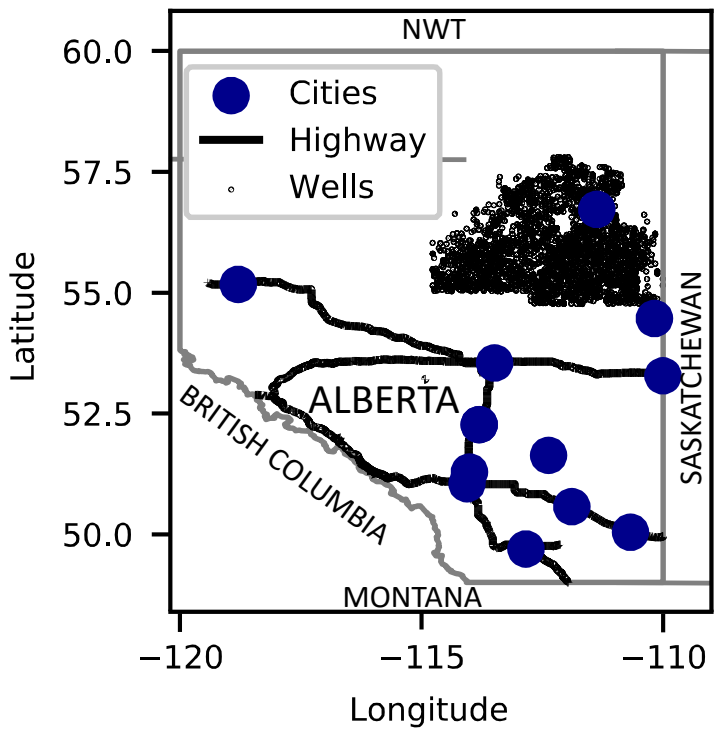

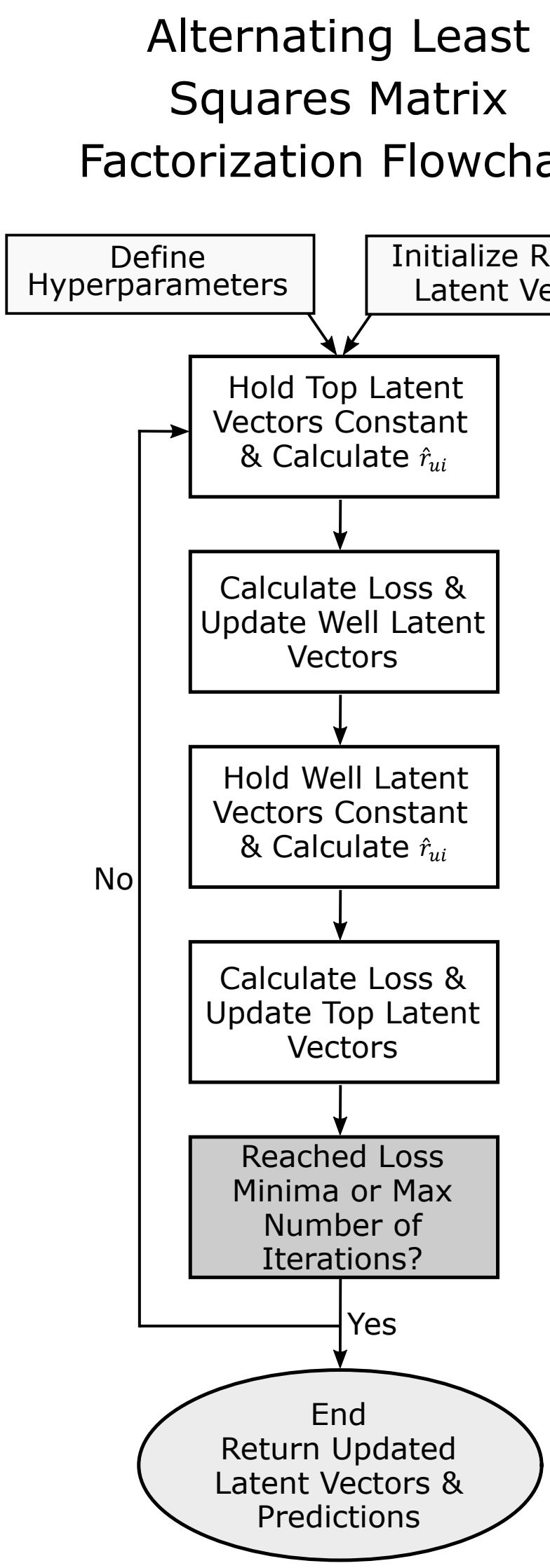

Alternating Least Squares Matrix Factorization Flowchart
Define Hyperparameters
Initialize Random Latent Vectors
Hold Top Latent Vectors Constant & Calculate $\hat{r}_{ui}$
Calculate Loss & Update Well Latent Vectors
Hold Well Latent Vectors Constant & Calculate $\hat{r}_{ui}$
Calculate Loss & Update Top Latent Vectors
Reached Loss Minima or Max Number of Iterations?
No
Yes
End
Return Updated Latent Vectors & Predictions

# Matrix Factorization Concept

| Formation Tops | Wells | | | | |
|---|---|---|---|---|---|
| | 3,347 | ? | ? | 3,149 | ? |
| | 7,108 | ? | 3,257 | 4,737 | ? |
| | ? | 5,557 | ? | 4,719 | ? |
| | 8,282 | ? | ? | ? | 4,976 |
| | ? | 8,587 | ? | ? | 1,284 |
| | ? | 9,595 | 6,050 | ? | ? |
| | 3,909 | 3,351 | ? | ? | ? |
| | 1,948 | ? | ? | 8,175 | 1,332 |
| | 8,846 | 7,751 | 9,805 | 5,535 | ? |
| | ? | 9,077 | ? | ? | 1,029 |

=

| Formation Tops Latent Vectors | | |
|---|---|---|
| | 0.76 | 0.70 |
| | 0.37 | 0.38 |
| | 0.74 | 0.52 |
| | 0.86 | 0.10 |
| | 0.66 | 0.88 |
| | 0.66 | 0.42 |
| | 0.53 | 0.43 |
| | 0.75 | 0.73 |
| | 0.45 | 0.71 |
| | 0.63 | 0.59 |

X

| Wells Latent Vectors | | | | |
|---|---|---|---|---|
| 0.07 | 0.21 | 0.66 | 0.80 | 0.86 |
| 0.76 | 0.47 | 0.27 | 0.12 | 0.46 |

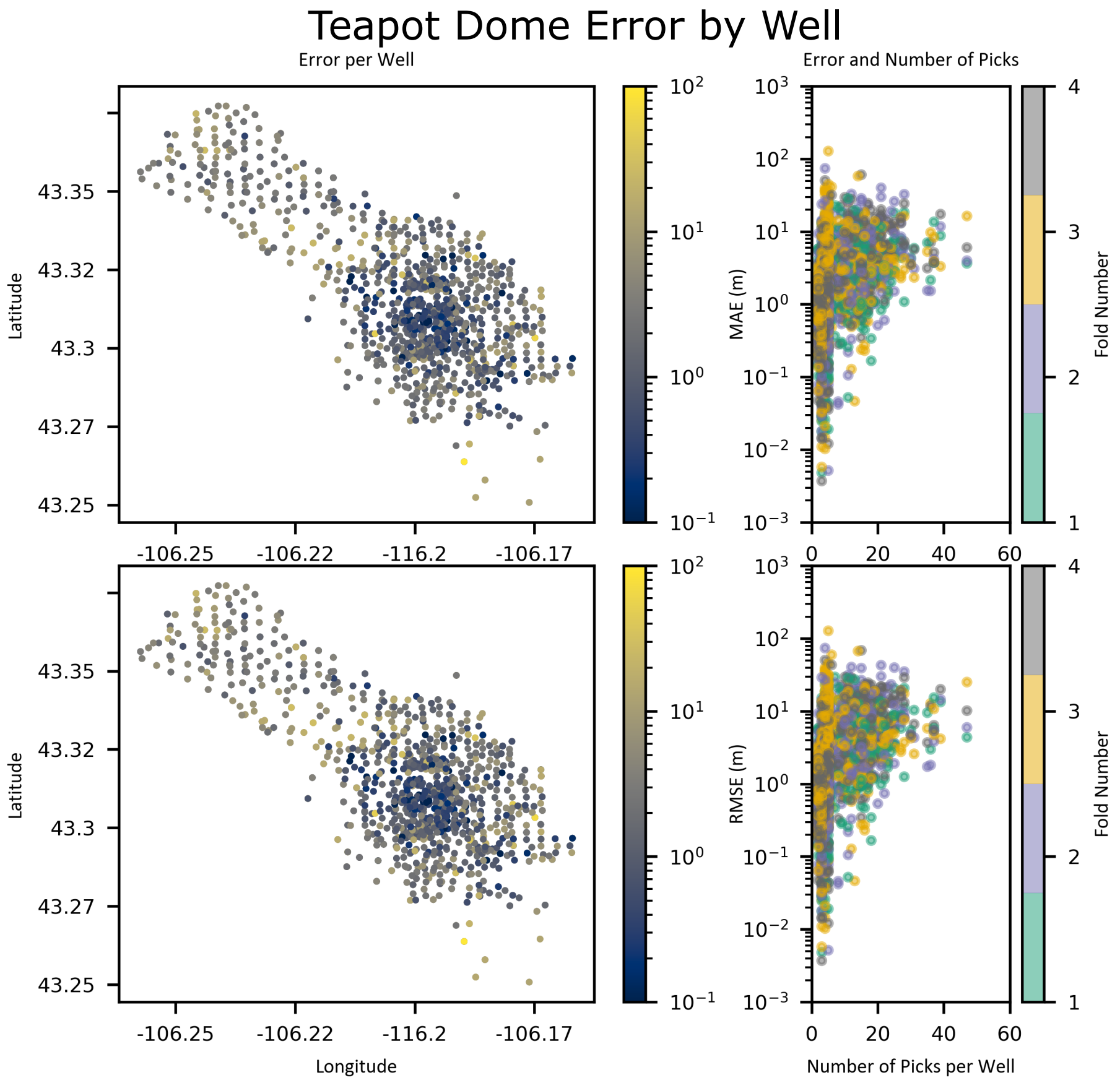

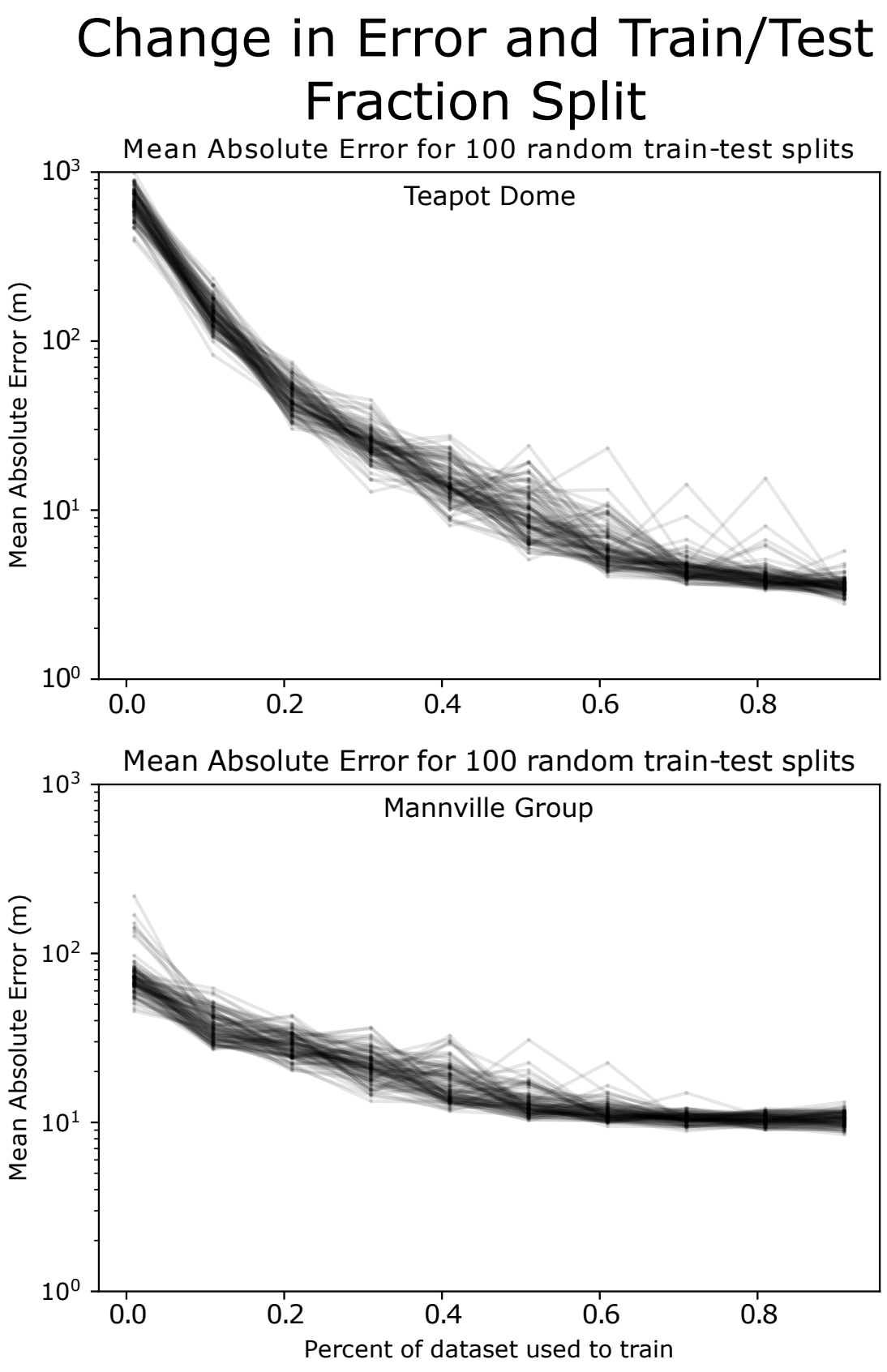
Change in Error and Train/Test Fraction Split
Mean Absolute Error for 100 random train-test splits
Teapot Dome
Mean Absolute Error (m)
10^3
10^2
10^1
10^0
0.0
0.2
0.4
0.6
0.8
Mean Absolute Error for 100 random train-test splits
Mannville Group
Mean Absolute Error (m)
10^3
10^2
10^1
10^0
0.0
0.2
0.4
0.6
0.8
Percent of dataset used to train

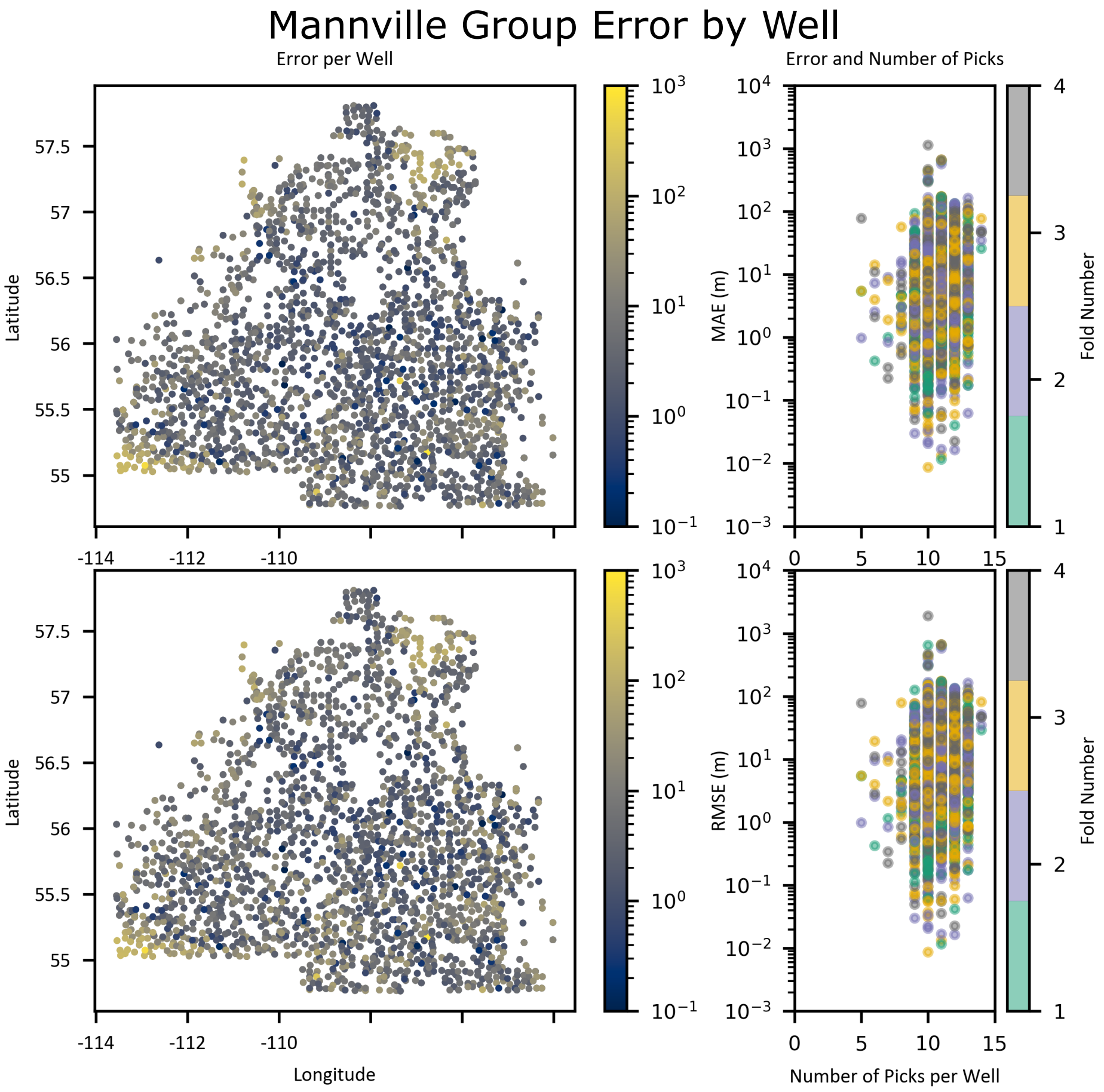

## Error by Tops

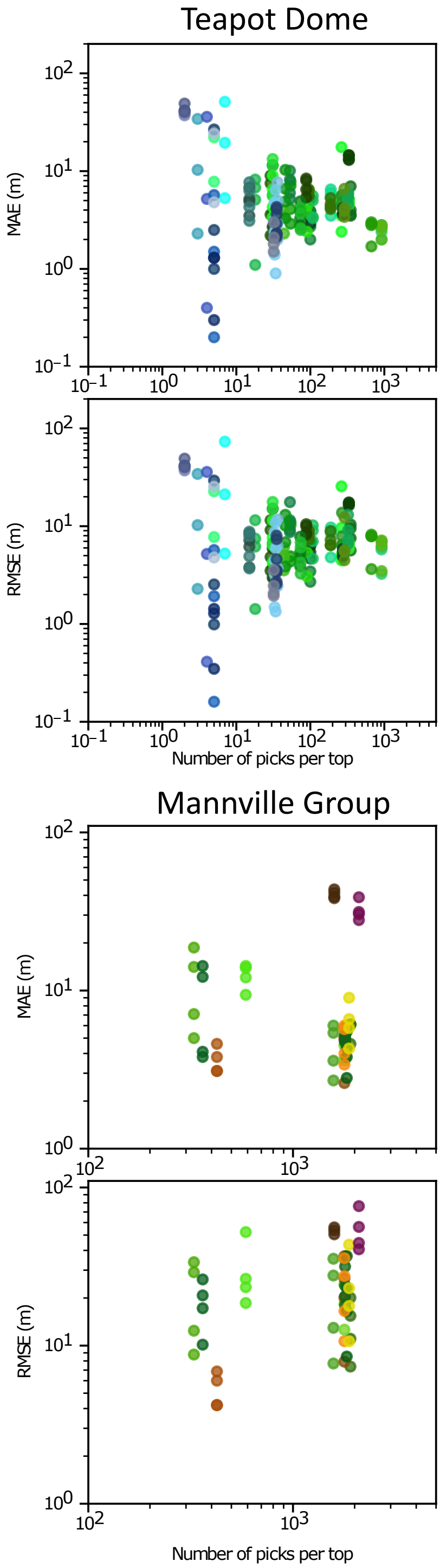

Table 1. Prediction Error for the Teapot Dome Dataset

| | Random Selection | |
|---|---|---|
| | Mean Absolute Error (m) | Root Mean Squared Error (m) |
| Fold 1 | 5.4 | 7.0 |
| Fold 2 | 6.9 | 9.5 |
| Fold 3 | 7.5 | 9.4 |
| Fold 4 | 7.4 | 9.7 |
| Average | 6.8 | 8.9 |
| | Spatially Blocked | |
| Fold 1 | 23.2 | 26.6 |
| Fold 2 | 27.7 | 27.7 |
| Fold 3 | 21.7 | 30.0 |
| Fold 4 | 39.7 | 23.7 |
| Average | 28.1 | 27.0 |

Table 2. Prediction Error for the Mannville Group Dataset

| | Mannville Group | |
|---|---|---|
| | Mean Absolute Error (m) | Root Mean Squared Error (m) |
| Fold 1 | 9.5 | 25.1 |
| Fold 2 | 11.8 | 24.5 |
| Fold 3 | 10.2 | 24.6 |
| Fold 4 | 10.9 | 35.3 |
| Average | 10.6 | 27.4 |
| | Spatially Blocked | |
| Fold 1 | 11.4 | 25.0 |
| Fold 2 | 13.2 | 24.1 |
| Fold 3 | 11.1 | 20.8 |
| Fold 4 | 11.2 | 40.5 |
| Average | 11.7 | 27.6 |

## Teapot Dome

### Recommender System

| Top Name | Fold 1 Number of Training Samples | Fold 1 Number of Holdout Samples | Fold 1 MAE (m) | Fold 1 RMSE (m) | Fold2 Number of Training Samples | Fold2 Number of Holdout Samples | Fold2 MAE (m) | Fold2 RMSE (m) | Fold 3 Number of Training Samples | Fold 3 Number of Holdout Samples | Fold 3 MAE (m) | Fold 3 RMSE (m) | Fold 4 Number of Training Samples | Fold 4 Number of Holdout Samples | Fold 4 MAE (m) | Fold 4 RMSE (m) | Average MAE (m) | Average RMSE (m) |
|---|---|---|---|---|---|---|---|---|---|---|---|---|---|---|---|---|---|---|
| SSXS | 198 | 66 | 2.4 | 4.5 | 202 | 62 | 4.8 | 9.6 | 198 | 66 | 17.4 | 25.4 | 197 | 67 | 3.5 | 5.1 | 7.0 | 11.1 |
| SSXbase | 280 | 71 | 3.5 | 5.6 | 260 | 91 | 3.7 | 7.6 | 264 | 87 | 6.9 | 9.5 | 254 | 97 | 4.0 | 10.4 | 4.5 | 8.3 |
| SHNNu | 707 | 209 | 2.0 | 3.3 | 720 | 196 | 2.4 | 6.1 | 682 | 234 | 2.8 | 7.0 | 707 | 209 | 2.8 | 5.8 | 2.5 | 5.6 |
| SHNNl | 707 | 213 | 2.0 | 3.5 | 718 | 202 | 2.7 | 6.3 | 704 | 216 | 2.6 | 6.9 | 696 | 224 | 2.8 | 7.1 | 2.5 | 6.0 |
| SHNNbs | 517 | 152 | 1.7 | 3.7 | 512 | 157 | 3.0 | 7.9 | 499 | 170 | 2.9 | 7.9 | 496 | 173 | 2.8 | 8.2 | 2.6 | 6.9 |
| StBR | 72 | 28 | 2.0 | 2.7 | 75 | 25 | 2.7 | 3.4 | 78 | 22 | 3.9 | 6.3 | 75 | 25 | 3.9 | 6.9 | 3.1 | 4.8 |
| StFT | 73 | 27 | 3.5 | 6.3 | 75 | 25 | 5.2 | 7.1 | 73 | 27 | 2.9 | 4.7 | 79 | 21 | 3.5 | 5.4 | 3.8 | 5.9 |
| StGD | 77 | 13 | 3.7 | 5.4 | 58 | 32 | 4.7 | 7.9 | 60 | 30 | 2.4 | 3.1 | 75 | 15 | 4.4 | 7.3 | 3.8 | 5.9 |
| StAM | 79 | 29 | 3.4 | 4.7 | 80 | 28 | 5.0 | 10.2 | 83 | 25 | 5.6 | 9.1 | 82 | 26 | 3.3 | 7.0 | 4.3 | 7.7 |
| NBRRws | 141 | 48 | 3.5 | 4.8 | 139 | 50 | 5.5 | 8.8 | 136 | 53 | 6.0 | 8.5 | 152 | 37 | 6.5 | 9.7 | 5.4 | 8.0 |
| NBRRsg | 140 | 48 | 4.1 | 6.4 | 147 | 41 | 4.3 | 7.2 | 139 | 49 | 4.5 | 6.4 | 138 | 50 | 5.3 | 9.0 | 4.6 | 7.2 |
| CRLL | 193 | 75 | 4.3 | 6.4 | 205 | 63 | 5.6 | 7.9 | 204 | 64 | 5.6 | 13.4 | 205 | 63 | 4.8 | 6.5 | 5.1 | 8.6 |
| F1WC | 245 | 87 | 4.6 | 5.9 | 235 | 97 | 5.2 | 6.5 | 256 | 76 | 5.1 | 6.4 | 260 | 72 | 4.8 | 6.7 | 4.9 | 6.4 |
| F1WCBench1Base | 70 | 25 | 5.5 | 7.3 | 74 | 21 | 6.5 | 9.0 | 71 | 24 | 5.7 | 7.9 | 71 | 24 | 6.4 | 8.5 | 6.0 | 8.2 |
| F1WCBench2Top | 66 | 22 | 8.4 | 10.0 | 77 | 11 | 5.4 | 8.2 | 65 | 23 | 6.5 | 8.5 | 56 | 32 | 8.0 | 10.5 | 7.1 | 9.3 |
| F1WCBench2Base | 52 | 21 | 3.9 | 5.4 | 53 | 20 | 3.3 | 4.8 | 56 | 17 | 4.9 | 7.8 | 58 | 15 | 2.8 | 3.9 | 3.7 | 5.5 |
| F1WCBench3Top | 56 | 18 | 4.2 | 6.1 | 54 | 20 | 2.6 | 4.3 | 60 | 14 | 2.5 | 3.3 | 52 | 22 | 3.7 | 6.8 | 3.3 | 5.1 |
| F1WCbase | 251 | 81 | 13.4 | 16.8 | 239 | 93 | 13.0 | 16.3 | 264 | 68 | 14.7 | 17.3 | 242 | 90 | 14.4 | 17.6 | 13.9 | 17.0 |
| B1 | 205 | 71 | 3.6 | 4.9 | 211 | 65 | 3.5 | 5.8 | 198 | 78 | 4.2 | 5.5 | 215 | 61 | 3.7 | 5.0 | 3.8 | 5.3 |
| F2WC | 239 | 85 | 5.8 | 9.1 | 246 | 78 | 6.0 | 13.0 | 247 | 77 | 4.6 | 6.6 | 243 | 81 | 5.2 | 7.3 | 5.4 | 9.0 |
| F2WCbase | 218 | 71 | 4.2 | 5.7 | 214 | 75 | 6.6 | 12.3 | 220 | 69 | 3.4 | 4.5 | 219 | 70 | 3.9 | 5.5 | 4.5 | 7.0 |
| F3WC | 37 | 17 | 2.9 | 4.4 | 42 | 12 | 7.2 | 10.4 | 43 | 11 | 4.1 | 4.9 | 40 | 14 | 7.8 | 11.0 | 5.5 | 7.7 |
| F3WCbase | 34 | 10 | 3.5 | 5.1 | 31 | 13 | 2.4 | 3.5 | 35 | 9 | 3.3 | 4.7 | 32 | 12 | 3.9 | 6.0 | 3.3 | 4.9 |
| MWRY | 38 | 15 | 6.0 | 9.2 | 41 | 12 | 8.3 | 11.7 | 37 | 16 | 5.2 | 8.7 | 43 | 10 | 10.0 | 17.5 | 7.4 | 11.8 |
| MDDY | 23 | 8 | 2.7 | 3.8 | 23 | 8 | 3.7 | 5.9 | 23 | 8 | 2.1 | 3.0 | 24 | 7 | 11.5 | 15.0 | 5.0 | 6.9 |
| THRM | 19 | 10 | 3.7 | 4.9 | 23 | 6 | 8.7 | 10.9 | 24 | 5 | 2.2 | 3.5 | 21 | 8 | 7.6 | 9.3 | 5.6 | 7.1 |
| DKOT | 34 | 12 | 5.0 | 6.8 | 33 | 13 | 6.8 | 10.4 | 34 | 12 | 3.4 | 3.9 | 37 | 9 | 10.5 | 13.3 | 6.4 | 8.6 |
| LKOT | 24 | 7 | 3.2 | 4.2 | 23 | 8 | 4.3 | 6.4 | 27 | 4 | 6.5 | 7.6 | 19 | 12 | 8.9 | 11.5 | 5.7 | 7.4 |
| MRSN | 22 | 9 | 5.5 | 6.6 | 26 | 5 | 11.4 | 15.2 | 23 | 8 | 5.6 | 6.6 | 22 | 9 | 13.4 | 17.7 | 9.0 | 11.5 |
| SNDCu | 4 | 3 | 19.7 | 21.3 | 4 | 3 | 50.4 | 72.0 | 7 | 0 | - | - | 6 | 1 | 5.2 | 5.2 | 25.1 | 32.9 |
| CNSP | 1 | 4 | 25.2 | 25.7 | 4 | 1 | 7.8 | 7.8 | 5 | 0 | - | - | 5 | 0 | - | - | 16.5 | 16.8 |
| CRMT | 13 | 5 | 5.1 | 6.2 | 13 | 5 | 1.1 | 1.4 | 13 | 5 | 6.7 | 7.5 | 15 | 3 | 8.2 | 11.5 | 5.3 | 6.6 |
| ALCV | 9 | 6 | 5.2 | 7.2 | 12 | 3 | 6.9 | 8.7 | 11 | 4 | 4.9 | 6.2 | 13 | 2 | 3.2 | 3.5 | 5.1 | 6.4 |
| RDPK | 12 | 3 | 3.1 | 3.7 | 11 | 4 | 6.3 | 7.9 | 10 | 5 | 7.7 | 8.8 | 12 | 3 | 4.3 | 4.9 | 5.4 | 6.3 |
| ERVY | 2 | 0 | - | - | 2 | 0 | - | - | 2 | 0 | - | - | 2 | 0 | - | - | - | - |
| FRLL | 2 | 1 | 36.7 | 36.7 | 3 | 0 | - | - | 2 | 1 | 2.3 | 2.3 | 2 | 1 | 10.2 | 10.2 | 16.4 | 16.4 |
| GLND | 1 | 0 | - | - | 1 | 0 | - | - | 1 | 0 | - | - | 1 | 0 | - | - | - | - |
| MNKT | 19 | 17 | 2.9 | 3.7 | 28 | 8 | 2.0 | 2.5 | 31 | 5 | 7.7 | 11.8 | 30 | 6 | 5.0 | 6.3 | 4.4 | 6.1 |
| OPCH | 24 | 10 | 2.1 | 2.4 | 25 | 9 | 0.9 | 1.3 | 29 | 5 | 6.4 | 11.0 | 24 | 10 | 4.5 | 5.9 | 3.5 | 5.2 |
| A Sand | 25 | 8 | 4.8 | 8.7 | 27 | 6 | 1.4 | 1.5 | 21 | 12 | 3.2 | 4.5 | 26 | 7 | 2.4 | 3.2 | 3.0 | 4.5 |
| B Dolo | 26 | 9 | 4.3 | 8.0 | 22 | 13 | 2.8 | 3.4 | 31 | 4 | 3.5 | 4.6 | 26 | 9 | 2.3 | 3.5 | 3.2 | 4.9 |
| B Sand | 24 | 11 | 4.1 | 7.4 | 30 | 5 | 2.3 | 2.6 | 28 | 7 | 2.3 | 2.7 | 23 | 12 | 2.0 | 3.5 | 2.7 | 4.0 |
| C1 Dolo | 27 | 5 | 1.4 | 2.0 | 22 | 10 | 1.8 | 2.1 | 23 | 9 | 3.0 | 3.6 | 24 | 8 | 2.1 | 2.5 | 2.1 | 2.5 |
| C1 Sand | 2 | 0 | - | - | 2 | 0 | - | - | 1 | 1 | 44.2 | 44.2 | 1 | 1 | 42.0 | 42.0 | 43.1 | 43.1 |
| C2 Dolo | 2 | 0 | - | - | 1 | 1 | - | - | 1 | 1 | 42.9 | 42.9 | 2 | 0 | - | - | 42.9 | 42.9 |
| C2 Sand | 5 | 0 | - | - | 4 | 1 | 2.5 | 2.5 | 4 | 1 | 1.1 | 1.1 | 2 | 3 | 26.7 | 29.3 | 10.1 | 11.0 |
| C3 Dolo | 3 | 1 | 0.4 | 0.4 | 2 | 2 | 5.2 | 5.2 | 4 | 0 | - | - | 3 | 1 | 35.9 | 35.9 | 13.8 | 13.8 |
| C3 Sand | 4 | 1 | 0.0 | 0.0 | 5 | 0 | - | - | 2 | 3 | 1.6 | 2.0 | 4 | 1 | 5.7 | 5.7 | 2.4 | 2.6 |
| C4 Dolo | 5 | 0 | - | - | 3 | 2 | 24.2 | 25.4 | 2 | 3 | 4.4 | 4.5 | 5 | 0 | - | - | 14.3 | 14.9 |
| C4 Sand | 4 | 1 | 0.4 | 0.4 | 4 | 1 | 0.0 | 0.0 | 3 | 2 | 1.3 | 1.4 | 4 | 1 | 1.3 | 1.3 | 0.8 | 0.8 |
| D Dolo | 2 | 0 | - | - | 1 | 1 | 40.4 | 40.4 | 1 | 1 | 52.5 | 52.5 | 2 | 0 | - | - | 46.5 | 46.4 |
| D Sand | 1 | 0 | - | - | 1 | 0 | - | - | 1 | 0 | - | - | 1 | 0 | - | - | - | - |
| E Dolo | 1 | 0 | - | - | 1 | 0 | - | - | 1 | 0 | - | - | 1 | 0 | - | - | - | - |
| AMSD | 1 | 0 | - | - | 1 | 0 | - | - | 1 | 0 | - | - | 1 | 0 | - | - | - | - |
| MDSN | 1 | 0 | - | - | 1 | 0 | - | - | 1 | 0 | - | - | 1 | 0 | - | - | - | - |
| PC | 1 | 0 | - | - | 1 | 0 | - | - | 1 | 0 | - | - | 1 | 0 | - | - | - | - |

### Green's Function Interpolation

| Top Name | Fold 1 MAE (m) | Fold 1 RMSE (m) | Fold 2 MAE (m) | Fold 2 RMSE (m) | Fold 3 MAE (m) | Fold 3 RMSE (m) | Fold 4 MAE (m) | Fold 4 RMSE (m) | Average MAE (m) | Average RMSE (m) |
|---|---|---|---|---|---|---|---|---|---|---|
| SSXS | 9.6 | 19.6 | 4.9 | 6.6 | 6.8 | 10.3 | 7.7 | 11.5 | 7.3 | 12.0 |
| SSXbase | 6.8 | 13.1 | 5.2 | 6.5 | 5.7 | 8.9 | 6.0 | 9.3 | 5.9 | 9.4 |
| SHNNu | 6.1 | 9.0 | 4.2 | 6.7 | 2.6 | 4.3 | 4.0 | 6.9 | 4.2 | 6.7 |
| SHNNl | 6.2 | 9.1 | 4.3 | 6.8 | 3.2 | 4.8 | 4.2 | 7.0 | 4.5 | 6.9 |
| SHNNbs | 6.6 | 11.6 | 5.1 | 8.7 | 3.5 | 5.1 | 4.5 | 7.6 | 4.9 | 8.2 |
| StBR | 49.9 | 182.1 | 9.3 | 16.4 | 7.5 | 13.6 | 7.3 | 12.4 | 18.5 | 56.1 |
| StFT | 49.7 | 181.4 | 8.9 | 15.5 | 8.1 | 13.1 | 6.0 | 11.2 | 18.2 | 55.3 |
| StGD | 40.6 | 155.7 | 7.7 | 11.3 | 6.8 | 9.1 | 5.4 | 7.5 | 15.1 | 45.9 |
| StAM | 42.5 | 188.6 | 6.5 | 9.7 | 7.4 | 10.4 | 9.6 | 20.9 | 16.5 | 57.4 |
| NBRRws | 7.1 | 13.3 | 7.4 | 9.9 | 6.3 | 8.5 | 7.0 | 15.9 | 6.9 | 11.9 |
| NBRRsg | 7.9 | 13.1 | 7.5 | 11.7 | 6.0 | 9.5 | 5.8 | 13.9 | 6.8 | 12.0 |
| CRLL | 4.4 | 6.6 | 6.1 | 9.5 | 4.8 | 9.0 | 5.2 | 8.3 | 5.1 | 8.4 |
| F1WC | 3.8 | 5.2 | 4.7 | 7.7 | 5.4 | 10.8 | 5.9 | 10.8 | 5.0 | 8.6 |
| F1WCBench1Base | 19.8 | 40.6 | 8.9 | 11.4 | 4.4 | 6.6 | 9.6 | 15.4 | 10.7 | 18.5 |
| F1WCBench2Top | 8.9 | 10.5 | 8.2 | 11.9 | 8.7 | 14.8 | 9.6 | 15.6 | 8.8 | 13.2 |
| F1WCBench2Base | 45.8 | 104.2 | 9.9 | 14.7 | 7.2 | 11.6 | 8.6 | 16.7 | 17.9 | 36.8 |
| F1WCBench3Top | 20.4 | 38.0 | 9.9 | 14.2 | 7.2 | 11.4 | 9.4 | 17.0 | 11.7 | 20.1 |
| F1WCbase | 9.2 | 13.5 | 11.0 | 15.6 | 11.6 | 18.5 | 9.4 | 15.5 | 10.3 | 15.8 |
| B1 | 3.3 | 4.5 | 4.1 | 5.7 | 6.3 | 12.4 | 7.0 | 13.8 | 5.2 | 9.1 |
| F2WC | 4.6 | 6.8 | 4.4 | 6.4 | 6.4 | 11.2 | 6.9 | 12.4 | 5.6 | 9.2 |
| F2WCbase | 4.0 | 5.3 | 5.9 | 8.1 | 5.1 | 8.9 | 7.2 | 14.5 | 5.6 | 9.2 |
| F3WC | 11.7 | 22.3 | 10.0 | 15.5 | 9.8 | 15.9 | 13.1 | 19.9 | 11.1 | 18.4 |
| F3WCbase | 6.6 | 7.9 | 12.3 | 18.5 | 7.4 | 10.6 | 9.5 | 12.8 | 8.9 | 12.4 |
| MWRY | 46.2 | 74.3 | 11.6 | 15.9 | 15.3 | 24.3 | 12.3 | 19.4 | 21.4 | 33.5 |
| MDDY | 49.0 | 63.2 | 14.6 | 17.4 | 17.8 | 21.7 | 13.4 | 20.6 | 23.7 | 30.7 |
| THRM | 47.7 | 60.2 | 15.9 | 18.3 | 14.9 | 17.9 | 13.3 | 21.7 | 23.0 | 29.5 |
| DKOT | 54.2 | 75.9 | 18.8 | 22.1 | 15.5 | 25.0 | 12.4 | 20.9 | 25.2 | 36.0 |
| LKOT | 26.0 | 39.3 | 17.6 | 29.3 | 10.5 | 15.1 | 18.8 | 26.9 | 18.2 | 27.7 |
| MRSN | 25.1 | 40.5 | 9.3 | 16.4 | 18.9 | 21.2 | 17.6 | 23.1 | 17.7 | 25.3 |
| SNDCu | 662.0 | 830.4 | 1669.1 | 2185.9 | 34.6 | 37.2 | 32.6 | 32.6 | 599.6 | 771.5 |
| CNSP | 120.4 | 136.8 | 134.0 | 134.0 | 0.8 | 0.8 | 2.6 | 2.6 | 64.5 | 68.6 |
| CRMT | 29.9 | 41.4 | 12.1 | 18.2 | 2.1 | 2.2 | 7.2 | 12.4 | 12.8 | 18.6 |
| ALCV | 34.6 | 41.0 | 17.6 | 21.1 | 11.7 | 20.3 | 0.9 | 1.1 | 16.2 | 20.9 |
| RDPK | 36.8 | 44.9 | 19.6 | 23.9 | 12.3 | 21.0 | 1.4 | 1.5 | 17.5 | 22.8 |
| ERVY | - | - | - | - | - | - | - | - | - | - |
| FRLL | - | - | - | - | - | - | - | - | - | - |
| GLND | - | - | - | - | - | - | - | - | - | - |
| MNKT | 17.0 | 25.1 | 16.2 | 21.3 | 7.1 | 8.7 | 2.0 | 3.0 | 10.6 | 14.5 |
| OPCH | 17.1 | 24.9 | 20.4 | 24.5 | 6.7 | 9.1 | 4.1 | 5.9 | 12.1 | 16.1 |
| A Sand | 17.7 | 24.3 | 15.7 | 21.3 | 8.3 | 11.0 | 4.7 | 6.2 | 11.6 | 15.7 |
| B Dolo | 23.3 | 41.6 | 15.7 | 21.5 | 8.5 | 10.8 | 3.1 | 3.7 | 12.6 | 19.4 |
| B Sand | 23.1 | 41.1 | 15.5 | 21.3 | 8.7 | 10.4 | 2.3 | 2.8 | 12.4 | 18.9 |
| C1 Dolo | 20.1 | 41.4 | 15.1 | 21.4 | 7.9 | 9.9 | 2.0 | 2.7 | 11.3 | 18.9 |
| C1 Sand | - | - | - | - | - | - | - | - | - | - |
| C2 Dolo | - | - | - | - | - | - | - | - | - | - |
| C2 Sand | 22.5 | 24.8 | 588.3 | 588.3 | 19.8 | 19.8 | 16.2 | 16.2 | 161.7 | 162.3 |
| C3 Dolo | 180.5 | 180.5 | 23.6 | 23.6 | 2.7 | 2.7 | 12.5 | 12.5 | 54.8 | 54.8 |
| C3 Sand | 23.1 | 24.4 | 760.6 | 760.6 | 18.4 | 18.4 | 20.3 | 20.3 | 205.6 | 205.9 |
| C4 Dolo | 24.5 | 26.0 | 846.8 | 846.8 | 20.6 | 20.6 | 20.5 | 20.5 | 228.1 | 228.5 |
| C4 Sand | 22.1 | 22.5 | 781.3 | 781.3 | 15.4 | 15.4 | 23.6 | 23.6 | 210.6 | 210.7 |
| D Dolo | - | - | - | - | - | - | - | - | - | - |
| D Sand | - | - | - | - | - | - | - | - | - | - |
| E Dolo | - | - | - | - | - | - | - | - | - | - |
| AMSD | - | - | - | - | - | - | - | - | - | - |
| MDSN | - | - | - | - | - | - | - | - | - | - |
| PC | - | - | - | - | - | - | - | - | - | - |

### Difference

| Top Name | Method MAE Difference* | Method RMSE Difference* |
|---|---|---|
| SSXS | 0.2 | 0.9 |
| SSXbase | 1.4 | 1.2 |
| SHNNu | 1.7 | 1.2 |
| SHNNl | 1.9 | 1.0 |
| SHNNbs | 2.3 | 1.3 |
| StBR | 15.4 | 51.3 |
| StFT | 14.4 | 49.4 |
| StGD | 11.3 | 40.0 |
| StAM | 12.2 | 49.7 |
| NBRRws | 1.5 | 3.9 |
| NBRRsg | 2.2 | 4.8 |
| CRLL | 0.1 | -0.2 |
| F1WC | 0.0 | 2.2 |
| F1WCBench1Base | 4.7 | 10.3 |
| F1WCBench2Top | 1.8 | 3.9 |
| F1WCBench2Base | 14.2 | 31.3 |
| F1WCBench3Top | 8.5 | 15.0 |
| F1WCbase | -3.6 | -1.2 |
| B1 | 1.4 | 3.7 |
| F2WC | 0.2 | 0.2 |
| F2WCbase | 1.0 | 2.2 |
| F3WC | 5.6 | 10.7 |
| F3WCbase | 5.7 | 7.6 |
| MWRY | 14.0 | 21.7 |
| MDDY | 18.7 | 23.8 |
| THRM | 17.4 | 22.4 |
| DKOT | 18.8 | 27.4 |
| LKOT | 12.5 | 20.2 |
| MRSN | 8.7 | 13.8 |
| SNDCu | 574.5 | 738.7 |
| CNSP | 48.0 | 51.8 |
| CRMT | 7.5 | 11.9 |
| ALCV | 11.1 | 14.5 |
| RDPK | 12.2 | 16.5 |
| ERVY | - | - |
| FRLL | - | - |
| GLND | - | - |
| MNKT | 6.2 | 8.5 |
| OPCH | 8.6 | 11.0 |
| A Sand | 8.7 | 11.2 |
| B Dolo | 9.4 | 14.5 |
| B Sand | 9.7 | 14.9 |
| C1 Dolo | 9.2 | 16.3 |
| C1 Sand | - | - |
| C2 Dolo | - | - |
| C2 Sand | 151.6 | 151.3 |
| C3 Dolo | 41.0 | 41.0 |
| C3 Sand | 203.2 | 203.4 |
| C4 Dolo | 213.8 | 213.5 |
| C4 Sand | 209.8 | 209.9 |
| D Dolo | - | - |
| D Sand | - | - |
| E Dolo | - | - |
| AMSD | - | - |
| MDSN | - | - |
| PC | - | - |

*Positive values mean RecSys outperformed state of the art Green's Function spline. Negative values mean Green's Function spline outperformed RecSys. Value is magnitude of outperformance in meters

## Mannville Group

### Recommender System

| Top Name | Fold 1 Number of Training Samples | Fold 1 Number of Holdout Samples | Fold 1 MAE (m) | Fold 1 RMSE (m) | Fold2 Number of Training Samples | Fold2 Number of Holdout Samples | Fold2 MAE (m) | Fold2 RMSE (m) | Fold 3 Number of Training Samples | Fold 3 Number of Holdout Samples | Fold 3 MAE (m) | Fold 3 RMSE (m) | Fold 4 Number of Training Samples | Fold 4 Number of Holdout Samples | Fold 4 MAE (m) | Fold 4 RMSE (m) | Average MAE (m) | Average RMSE (m) |
|---|---|---|---|---|---|---|---|---|---|---|---|---|---|---|---|---|---|---|
| Mannville Group | 1196 | 395 | 38.4 | 50.6 | 1189 | 402 | 39.2 | 55.9 | 1195 | 396 | 41.5 | 53.5 | 1195 | 396 | 43.6 | 170.7 | 40.7 | 82.7 |
| Transgression 8 | 245 | 83 | 5.0 | 8.8 | 242 | 86 | 18.7 | 33.7 | 248 | 80 | 7.1 | 12.4 | 250 | 78 | 14.1 | 29.1 | 11.2 | 21.0 |
| Transgression 7 | 275 | 87 | 4.1 | 17.2 | 254 | 108 | 14.3 | 26.2 | 284 | 78 | 3.8 | 10.1 | 274 | 88 | 12.2 | 20.8 | 8.6 | 18.6 |
| Transgression 6 | 438 | 150 | 14.3 | 52.2 | 447 | 141 | 13.9 | 26.4 | 434 | 154 | 12.1 | 23.4 | 445 | 143 | 9.4 | 18.6 | 12.4 | 30.2 |
| Transgression 5 | 1410 | 498 | 6.1 | 15.4 | 1425 | 483 | 6.1 | 11.1 | 1409 | 499 | 6.1 | 20.1 | 1481 | 427 | 4.6 | 7.4 | 5.7 | 13.5 |
| Clearwater/Wabiskaw | 320 | 105 | 3.1 | 4.2 | 319 | 106 | 3.8 | 6.0 | 324 | 101 | 3.1 | 4.2 | 312 | 113 | 4.6 | 6.9 | 3.7 | 5.3 |
| Transgression 4 | 1171 | 405 | 6.0 | 27.8 | 1221 | 355 | 2.7 | 7.7 | 1195 | 381 | 3.6 | 13.0 | 1141 | 435 | 5.4 | 35.4 | 4.4 | 21.0 |
| Regression 3 | 0 | 0 | - | - | 0 | 0 | - | - | 0 | 0 | - | - | 0 | 0 | - | - | - | - |
| Transgression 3 | 1345 | 438 | 3.7 | 12.6 | 1356 | 427 | 3.6 | 18.4 | 1336 | 447 | 5.7 | 35.1 | 1313 | 470 | 4.5 | 24.4 | 4.4 | 22.6 |
| Regression 2 | 1338 | 445 | 2.6 | 7.9 | 1346 | 437 | 5.1 | 26.6 | 1342 | 441 | 4.9 | 20.1 | 1323 | 460 | 5.8 | 36.8 | 4.6 | 22.8 |
| Transgression 2 | 1382 | 449 | 5.5 | 36.8 | 1380 | 451 | 3.8 | 16.5 | 1349 | 482 | 4.3 | 23.2 | 1383 | 448 | 2.8 | 8.5 | 4.1 | 21.3 |
| Transgression 1 | 1358 | 439 | 4.8 | 20.5 | 1336 | 461 | 5.2 | 24.0 | 1380 | 417 | 5.0 | 18.0 | 1318 | 479 | 4.7 | 31.6 | 4.9 | 23.6 |
| Regression 1 | 1336 | 447 | 3.4 | 10.7 | 1330 | 453 | 4.0 | 16.6 | 1321 | 462 | 5.7 | 27.5 | 1363 | 420 | 6.0 | 35.7 | 4.8 | 22.6 |
| McMurray Formation | 1401 | 479 | 4.3 | 10.6 | 1393 | 487 | 5.8 | 17.8 | 1437 | 443 | 9.0 | 43.5 | 1425 | 455 | 6.6 | 23.1 | 6.4 | 23.8 |
| Paleozoic | 1595 | 504 | 31.4 | 76.2 | 1567 | 532 | 38.9 | 56.3 | 1559 | 540 | 30.4 | 40.6 | 1594 | 505 | 27.9 | 44.6 | 32.2 | 54.4 |

### Green's Function Interpolation

| Top Name | Fold 1 MAE (m) | Fold 1 RMSE (m) | Fold 2 MAE (m) | Fold 2 RMSE (m) | Fold 3 MAE (m) | Fold 3 RMSE (m) | Fold 4 MAE (m) | Fold 4 RMSE (m) | Average MAE (m) | Average RMSE (m) |
|---|---|---|---|---|---|---|---|---|---|---|
| Mannville Group | 27.9 | 309.6 | 13.2 | 41.7 | 13.5 | 49.8 | 47.1 | 91.1 | 25.4 | 123.0 |
| Transgression 8 | 16.4 | 50.5 | 22.0 | 43.9 | 18.4 | 49.1 | 14.2 | 29.9 | 17.8 | 43.3 |
| Transgression 7 | 49.9 | 97.6 | 17.4 | 60.6 | 13.7 | 48.8 | 25.4 | 50.5 | 26.6 | 64.4 |
| Transgression 6 | 19.3 | 55.5 | 10.2 | 12.7 | 8.5 | 10.5 | 109.3 | 212.8 | 36.8 | 72.9 |
| Transgression 5 | 26.1 | 281.6 | 8.9 | 33.1 | 19.5 | 51.1 | 34.9 | 75.4 | 22.4 | 110.3 |
| Clearwater/Wabiskaw | 8.2 | 13.8 | 6.3 | 8.2 | 6.4 | 12.1 | 8.7 | 13.5 | 7.4 | 11.9 |
| Transgression 4 | 29.8 | 311.2 | 14.5 | 42.1 | 20.5 | 57.5 | 29.0 | 70.7 | 23.4 | 120.4 |
| Regression 3 | - | - | - | - | - | - | - | - | - | - |
| Transgression 3 | 28.8 | 292.6 | 9.8 | 34.4 | 9.4 | 14.8 | 18.6 | 60.4 | 16.6 | 100.6 |
| Regression 2 | 28.7 | 292.6 | 9.6 | 34.3 | 9.4 | 14.8 | 18.4 | 60.3 | 16.5 | 100.5 |
| Transgression 2 | 28.3 | 288.8 | 11.7 | 36.0 | 13.1 | 39.5 | 54.1 | 93.9 | 26.8 | 114.5 |
| Transgression 1 | 29.0 | 291.1 | 12.5 | 36.6 | 14.1 | 39.7 | 116.7 | 203.3 | 43.1 | 142.7 |
| Regression 1 | 29.2 | 292.4 | 12.0 | 36.5 | 14.4 | 40.0 | 214.3 | 351.3 | 67.5 | 180.0 |
| McMurray Formation | 28.3 | 284.9 | 11.4 | 35.6 | 14.8 | 39.6 | 22.8 | 58.4 | 19.3 | 104.6 |
| Paleozoic | 33.8 | 270.4 | 16.6 | 36.2 | 18.9 | 41.6 | 30.7 | 64.3 | 25.0 | 103.1 |

### Difference

| Top Name | Method MAE Difference* | Method RMSE Difference* |
|---|---|---|
| Mannville Group | -15.2 | 40.4 |
| Transgression 8 | 6.5 | 22.3 |
| Transgression 7 | 18.0 | 45.8 |
| Transgression 6 | 24.4 | 42.7 |
| Transgression 5 | 16.6 | 96.8 |
| Clearwater/Wabiskaw | 3.7 | 6.6 |
| Transgression 4 | 19.0 | 99.4 |
| Regression 3 | - | - |
| Transgression 3 | 12.3 | 77.9 |
| Regression 2 | 11.9 | 77.6 |
| Transgression 2 | 22.7 | 93.3 |
| Transgression 1 | 38.2 | 119.1 |
| Regression 1 | 62.7 | 157.4 |
| McMurray Formation | 12.9 | 80.9 |
| Paleozoic | -7.2 | 48.7 |

## Teapot Dome

### Recommender System

| Top Name | Fold 1 Number of Training Samples | Fold 1 Number of Holdout Samples | Fold 1 MAE (m) | Fold 1 RMSE (m) | Fold2 Number of Training Samples | Fold2 Number of Holdout Samples | Fold2 MAE (m) | Fold2 RMSE (m) | Fold 3 Number of Training Samples | Fold 3 Number of Holdout Samples | Fold 3 MAE (m) | Fold 3 RMSE (m) | Fold 4 Number of Training Samples | Fold 4 Number of Holdout Samples | Fold 4 MAE (m) | Fold 4 RMSE (m) | Average MAE (m) | Average RMSE (m) |
|---|---|---|---|---|---|---|---|---|---|---|---|---|---|---|---|---|---|---|
| SSXS | 201 | 63 | 1.5 | 2.5 | 189 | 75 | 1.9 | 3.9 | 202 | 62 | 1.5 | 2.9 | 200 | 64 | 2.1 | 3.2 | 1.8 | 3.1 |
| SSXbase | 262 | 89 | 2.7 | 4.5 | 254 | 97 | 1.8 | 2.7 | 267 | 84 | 2.0 | 3.8 | 270 | 81 | 2.2 | 4.0 | 2.2 | 3.8 |
| SHNNu | 673 | 243 | 1.9 | 3.2 | 684 | 232 | 2.0 | 2.9 | 691 | 225 | 3.5 | 8.3 | 700 | 216 | 2.0 | 3.0 | 2.3 | 4.3 |
| SHNNl | 679 | 241 | 2.9 | 5.7 | 685 | 235 | 1.6 | 2.4 | 694 | 226 | 2.7 | 8.3 | 702 | 218 | 1.8 | 2.9 | 2.2 | 4.8 |
| SHNNbs | 499 | 170 | 2.2 | 6.6 | 498 | 171 | 1.7 | 3.6 | 501 | 168 | 1.7 | 2.9 | 509 | 160 | 2.1 | 3.8 | 1.9 | 4.2 |
| StBR | 77 | 23 | 2.3 | 3.2 | 73 | 27 | 3.2 | 5.8 | 76 | 24 | 3.0 | 5.2 | 74 | 26 | 2.1 | 2.9 | 2.6 | 4.3 |
| StFT | 78 | 22 | 3.3 | 5.0 | 72 | 28 | 3.6 | 5.9 | 76 | 24 | 4.3 | 6.8 | 74 | 26 | 2.3 | 3.3 | 3.4 | 5.3 |
| StGD | 70 | 20 | 3.6 | 5.1 | 68 | 22 | 2.1 | 3.3 | 67 | 23 | 5.8 | 10.5 | 65 | 25 | 3.7 | 5.0 | 3.8 | 6.0 |
| StAM | 86 | 22 | 3.7 | 5.1 | 80 | 28 | 5.5 | 17.8 | 81 | 27 | 4.5 | 8.1 | 77 | 31 | 5.3 | 8.5 | 4.7 | 9.9 |
| NBRRws | 147 | 42 | 5.2 | 7.6 | 142 | 47 | 4.9 | 7.3 | 139 | 50 | 4.9 | 7.7 | 139 | 50 | 5.4 | 8.1 | 5.1 | 7.7 |
| NBRRsg | 145 | 43 | 5.6 | 8.4 | 144 | 44 | 2.9 | 3.8 | 137 | 51 | 4.3 | 7.9 | 138 | 50 | 4.6 | 7.0 | 4.4 | 6.8 |
| CRLL | 212 | 56 | 5.7 | 8.5 | 201 | 67 | 4.0 | 5.2 | 194 | 74 | 7.1 | 12.8 | 197 | 71 | 7.2 | 10.9 | 6.0 | 9.3 |
| F1WC | 255 | 77 | 5.1 | 6.4 | 249 | 83 | 5.0 | 6.6 | 245 | 87 | 4.5 | 5.4 | 247 | 85 | 5.0 | 6.6 | 4.9 | 6.2 |
| F1WCBench1Base | 69 | 26 | 5.2 | 6.6 | 65 | 30 | 5.5 | 7.4 | 77 | 18 | 4.8 | 6.5 | 74 | 21 | 7.9 | 10.0 | 5.9 | 7.6 |
| F1WCBench2Top | 63 | 25 | 6.8 | 8.6 | 62 | 26 | 7.3 | 9.6 | 73 | 15 | 8.0 | 10.1 | 66 | 22 | 6.5 | 8.4 | 7.1 | 9.2 |
| F1WCBench2Base | 50 | 23 | 2.8 | 4.0 | 52 | 21 | 3.3 | 4.9 | 61 | 12 | 3.2 | 6.3 | 56 | 17 | 4.2 | 5.1 | 3.4 | 5.1 |
| F1WCBench3Top | 53 | 21 | 2.2 | 3.1 | 54 | 20 | 2.2 | 3.7 | 59 | 15 | 1.7 | 2.4 | 56 | 18 | 2.5 | 3.6 | 2.1 | 3.2 |
| F1WCbase | 252 | 80 | 12.7 | 14.9 | 250 | 82 | 12.7 | 15.2 | 245 | 87 | 14.2 | 16.8 | 249 | 83 | 10.9 | 13.5 | 12.6 | 15.1 |
| B1 | 210 | 66 | 5.4 | 9.2 | 209 | 67 | 3.4 | 4.5 | 204 | 72 | 3.2 | 4.0 | 205 | 71 | 3.8 | 5.0 | 3.9 | 5.7 |
| F2WC | 249 | 75 | 4.8 | 6.4 | 244 | 80 | 5.2 | 8.5 | 239 | 85 | 3.9 | 6.3 | 240 | 84 | 4.8 | 7.4 | 4.7 | 7.2 |
| F2WCbase | 221 | 68 | 3.7 | 5.7 | 218 | 71 | 3.4 | 4.3 | 216 | 73 | 3.1 | 4.0 | 212 | 77 | 3.7 | 5.2 | 3.5 | 4.8 |
| F3WC | 40 | 14 | 2.8 | 3.7 | 43 | 11 | 3.5 | 4.3 | 40 | 14 | 2.5 | 3.2 | 39 | 15 | 2.7 | 3.7 | 2.9 | 3.7 |
| F3WCbase | 33 | 11 | 3.6 | 4.7 | 35 | 9 | 2.5 | 3.2 | 33 | 11 | 2.2 | 2.9 | 31 | 13 | 2.0 | 2.4 | 2.6 | 3.3 |
| MWRY | 39 | 14 | 16.9 | 38.0 | 43 | 10 | 5.0 | 5.8 | 37 | 16 | 4.0 | 8.2 | 40 | 13 | 2.6 | 3.3 | 7.1 | 13.8 |
| MDDY | 24 | 7 | 4.5 | 5.7 | 28 | 3 | 2.3 | 2.3 | 23 | 8 | 2.0 | 2.6 | 18 | 13 | 3.7 | 6.1 | 3.1 | 4.2 |
| THRM | 22 | 7 | 4.6 | 6.2 | 26 | 3 | 3.1 | 3.2 | 23 | 6 | 2.2 | 2.9 | 16 | 13 | 4.2 | 6.3 | 3.5 | 4.7 |
| DKOT | 37 | 9 | 7.3 | 9.5 | 36 | 10 | 5.1 | 5.5 | 33 | 13 | 3.0 | 5.4 | 32 | 14 | 4.9 | 6.5 | 5.1 | 6.7 |
| LKOT | 24 | 7 | 6.3 | 8.7 | 26 | 5 | 3.1 | 3.8 | 20 | 11 | 5.1 | 6.0 | 23 | 8 | 4.8 | 6.6 | 4.8 | 6.3 |
| MRSN | 25 | 6 | 7.2 | 10.3 | 25 | 6 | 5.9 | 7.7 | 20 | 11 | 7.3 | 8.3 | 23 | 8 | 5.7 | 7.6 | 6.5 | 8.5 |
| SNDCu | 3 | 4 | 10.3 | 11.1 | 7 | 0 | - | - | 5 | 2 | 8.6 | 9.1 | 6 | 1 | 7.0 | 7.0 | 8.6 | 9.1 |
| CNSP | 3 | 2 | 20.3 | 20.8 | 5 | 0 | - | - | 4 | 1 | 12.9 | 12.9 | 3 | 2 | 8.4 | 9.0 | 13.9 | 14.2 |
| CRMT | 13 | 5 | 17.9 | 31.0 | 13 | 5 | 8.4 | 12.6 | 14 | 4 | 1.1 | 1.2 | 14 | 4 | 23.1 | 41.8 | 12.6 | 21.6 |
| ALCV | 12 | 3 | 6.9 | 10.4 | 10 | 5 | 12.2 | 16.3 | 11 | 4 | 2.9 | 3.2 | 12 | 3 | 7.2 | 9.7 | 7.3 | 9.9 |
| RDPK | 12 | 3 | 6.3 | 8.5 | 10 | 5 | 11.3 | 16.5 | 11 | 4 | 3.0 | 3.2 | 12 | 3 | 7.2 | 10.0 | 6.9 | 9.6 |
| ERVY | - | - | - | - | - | - | - | - | - | - | - | - | - | - | - | - | - | - |
| FRLL | - | - | - | - | - | - | - | - | - | - | - | - | - | - | - | - | - | - |
| GLND | - | - | - | - | - | - | - | - | - | - | - | - | - | - | - | - | - | - |
| MNKT | 26 | 10 | 52.2 | 64.5 | 28 | 8 | 66.0 | 75.7 | 27 | 9 | 46.7 | 51.0 | 27 | 9 | 93.0 | 120.8 | 64.5 | 78.0 |
| OPCH | 24 | 10 | 58.2 | 69.4 | 26 | 8 | 72.3 | 81.1 | 25 | 9 | 53.0 | 57.0 | 27 | 7 | 119.0 | 141.5 | 75.6 | 87.2 |
| A Sand | 24 | 9 | 78.4 | 88.1 | 25 | 8 | 90.8 | 97.6 | 25 | 8 | 71.0 | 74.4 | 25 | 8 | 124.9 | 148.0 | 91.3 | 102.0 |
| B Dolo | 25 | 10 | 83.8 | 92.2 | 28 | 7 | 90.4 | 95.2 | 26 | 9 | 78.0 | 81.0 | 26 | 9 | 125.4 | 147.4 | 94.4 | 103.9 |
| B Sand | 25 | 10 | 90.3 | 98.0 | 28 | 7 | 97.0 | 101.5 | 26 | 9 | 84.3 | 87.1 | 26 | 9 | 132.2 | 153.2 | 100.9 | 109.9 |
| C1 Dolo | 23 | 9 | 110.7 | 117.5 | 26 | 6 | 104.5 | 105.5 | 23 | 9 | 102.7 | 105.0 | 24 | 8 | 135.2 | 151.0 | 113.3 | 119.7 |
| C1 Sand | - | - | - | - | - | - | - | - | - | - | - | - | - | - | - | - | - | - |
| C2 Dolo | - | - | - | - | - | - | - | - | - | - | - | - | - | - | - | - | - | - |
| C2 Sand | 4 | 1 | 66.7 | 66.7 | 4 | 1 | 98.7 | 98.7 | 4 | 1 | 73.7 | 73.7 | 3 | 2 | 165.5 | 167.2 | 101.1 | 101.6 |
| C3 Dolo | 3 | 1 | 70.1 | 70.1 | 3 | 1 | 100.3 | 100.3 | 3 | 1 | 75.6 | 75.6 | 3 | 1 | 194.1 | 194.1 | 110.0 | 110.0 |
| C3 Sand | 4 | 1 | 73.5 | 73.5 | 4 | 1 | 105.6 | 105.6 | 4 | 1 | 78.6 | 78.6 | 3 | 2 | 171.7 | 173.8 | 107.3 | 107.9 |
| C4 Dolo | 4 | 1 | 77.2 | 77.2 | 4 | 1 | 107.3 | 107.3 | 4 | 1 | 81.4 | 81.4 | 3 | 2 | 175.4 | 177.4 | 110.3 | 110.8 |
| C4 Sand | 4 | 1 | 80.6 | 80.6 | 4 | 1 | 112.2 | 112.2 | 4 | 1 | 84.1 | 84.1 | 3 | 2 | 177.5 | 179.5 | 113.6 | 114.1 |
| D Dolo | 2 | 0 | - | - | 1 | 1 | - | - | 1 | 1 | - | - | 2 | 0 | - | - | - | - |
| D Sand | 1 | 0 | - | - | 1 | 0 | - | - | 1 | 0 | - | - | 1 | 0 | - | - | - | - |
| E Dolo | 1 | 0 | - | - | 1 | 0 | - | - | 1 | 0 | - | - | 1 | 0 | - | - | - | - |
| AMSD | 1 | 0 | - | - | 1 | 0 | - | - | 1 | 0 | - | - | 1 | 0 | - | - | - | - |
| MDSN | 1 | 0 | - | - | 1 | 0 | - | - | 1 | 0 | - | - | 1 | 0 | - | - | - | - |
| PC | 1 | 0 | - | - | 1 | 0 | - | - | 1 | 0 | - | - | 1 | 0 | - | - | - | - |

### Green's Function Interpolation

| Top Name | Fold 1 MAE (m) | Fold 1 RMSE (m) | Fold 2 MAE (m) | Fold 2 RMSE (m) | Fold 3 MAE (m) | Fold 3 RMSE (m) | Fold 4 MAE (m) | Fold 4 RMSE (m) | Average MAE (m) | Average RMSE (m) | Difference: Method MAE Difference* | Difference: Method RMSE Difference* |
|---|---|---|---|---|---|---|---|---|---|---|---|---|
| SSXS | 5.9 | 9.0 | 5.1 | 7.4 | 4.3 | 5.5 | 5.2 | 6.9 | 5.1 | 7.2 | 3.4 | 4.1 |
| SSXbase | 3.8 | 5.1 | 4.6 | 7.1 | 4.3 | 6.2 | 4.8 | 9.2 | 4.4 | 6.9 | 2.2 | 3.2 |
| SHNNu | 4.2 | 7.8 | 3.6 | 7.2 | 2.7 | 4.4 | 3.5 | 5.8 | 3.5 | 6.3 | 1.2 | 2.0 |
| SHNNl | 2.8 | 4.8 | 2.7 | 4.7 | 3.8 | 7.3 | 3.1 | 5.1 | 3.1 | 5.5 | 0.9 | 0.7 |
| SHNNbs | 3.7 | 5.1 | 3.2 | 4.6 | 4.1 | 7.3 | 4.1 | 7.6 | 3.8 | 6.1 | 1.8 | 1.9 |
| StBR | 103.1 | 310.3 | 5.7 | 7.9 | 107.4 | 325.3 | 5.0 | 6.7 | 55.3 | 162.6 | 52.7 | 158.3 |
| StFT | 92.2 | 282.4 | 2.1 | 2.3 | 3.7 | 4.9 | 5.1 | 7.9 | 25.7 | 74.4 | 22.4 | 69.1 |
| StGD | 94.5 | 271.7 | 1.9 | 2.4 | 6.8 | 13.1 | 90.4 | 260.2 | 48.4 | 136.8 | 44.6 | 130.9 |
| StAM | 90.8 | 294.2 | 5.2 | 8.9 | 3.7 | 5.6 | 17.0 | 38.0 | 29.2 | 86.7 | 24.4 | 76.9 |
| NBRRws | 3.4 | 4.3 | 4.1 | 6.0 | 6.9 | 9.3 | 5.9 | 8.4 | 5.1 | 7.0 | 0.0 | -0.7 |
| NBRRsg | 5.7 | 10.2 | 4.7 | 6.6 | 5.4 | 6.9 | 10.4 | 22.8 | 6.5 | 11.6 | 2.1 | 4.8 |
| CRLL | 3.0 | 3.9 | 4.2 | 8.4 | 5.5 | 9.9 | 6.1 | 12.0 | 4.7 | 8.5 | -1.3 | -0.8 |
| F1WC | 4.4 | 7.9 | 2.9 | 4.6 | 3.7 | 5.9 | 4.3 | 7.0 | 3.8 | 6.4 | -1.1 | 0.1 |
| F1WCBench1Base | 5.1 | 7.2 | 5.6 | 7.9 | 2.5 | 3.7 | 8.2 | 11.6 | 5.4 | 7.6 | -0.5 | 0.0 |
| F1WCBench2Top | 8.3 | 13.7 | 8.2 | 11.4 | 7.6 | 10.8 | 11.7 | 14.9 | 8.9 | 12.7 | 1.8 | 3.5 |
| F1WCBench2Base | 5.6 | 7.0 | 9.0 | 11.3 | 4.2 | 7.8 | 8.7 | 11.6 | 6.9 | 9.4 | 3.5 | 4.4 |
| F1WCBench3Top | 7.3 | 9.2 | 4.6 | 6.9 | 6.2 | 8.0 | 8.9 | 11.7 | 6.8 | 9.0 | 4.6 | 5.8 |
| F1WCbase | 10.0 | 15.0 | 8.7 | 12.6 | 7.2 | 10.3 | 7.8 | 11.9 | 8.4 | 12.5 | -4.2 | -2.7 |
| B1 | 3.5 | 5.5 | 4.6 | 7.7 | 2.9 | 4.0 | 4.2 | 6.5 | 3.8 | 5.9 | -0.1 | 0.3 |
| F2WC | 6.8 | 14.4 | 4.6 | 7.0 | 7.1 | 11.9 | 5.1 | 9.7 | 5.9 | 10.8 | 1.2 | 3.6 |
| F2WCbase | 6.5 | 9.4 | 5.0 | 7.1 | 3.5 | 4.4 | 4.9 | 10.0 | 5.0 | 7.7 | 1.5 | 2.9 |
| F3WC | 8.2 | 14.8 | 8.9 | 12.5 | 12.0 | 20.7 | 22.8 | 28.0 | 13.0 | 19.0 | 10.1 | 15.3 |
| F3WCbase | 3.6 | 4.1 | 7.3 | 8.5 | 3.8 | 4.7 | 8.4 | 9.2 | 5.7 | 6.6 | 3.2 | 3.3 |
| MWRY | 19.2 | 30.8 | 25.6 | 35.1 | 11.0 | 23.1 | 21.6 | 29.6 | 19.3 | 29.6 | 12.2 | 15.8 |
| MDDY | 33.4 | 52.9 | 39.8 | 52.0 | 42.3 | 64.5 | 40.5 | 64.0 | 39.0 | 58.3 | 35.9 | 54.1 |
| THRM | 16.1 | 17.6 | 13.6 | 16.7 | 14.5 | 17.6 | 8.2 | 9.9 | 13.1 | 15.5 | 9.6 | 10.8 |
| DKOT | 16.2 | 18.9 | 5.6 | 6.2 | 4.6 | 5.7 | 15.1 | 25.5 | 10.4 | 14.1 | 5.3 | 7.4 |
| LKOT | 8.8 | 10.4 | 12.9 | 13.0 | 8.3 | 10.0 | 9.5 | 10.1 | 9.9 | 10.9 | 5.0 | 4.6 |
| MRSN | 13.9 | 15.0 | 13.2 | 15.2 | 25.7 | 39.3 | 8.4 | 9.7 | 15.3 | 19.8 | 8.8 | 11.3 |
| SNDCu | 58.7 | 58.7 | 223.8 | 223.8 | 223.8 | 223.8 | 11.9 | 11.9 | 129.5 | 129.5 | 120.9 | 120.5 |
| CNSP | 0.8 | 0.8 | 34.4 | 34.4 | 134.0 | 134.0 | 2.6 | 2.6 | 43.0 | 43.0 | 29.1 | 28.8 |
| CRMT | 45.3 | 54.3 | 26.0 | 26.5 | 3.7 | 4.5 | 16.3 | 20.6 | 22.8 | 26.5 | 10.2 | 4.8 |
| ALCV | 40.6 | 40.8 | 2.3 | 2.3 | 4.5 | 5.4 | 30.8 | 42.9 | 19.5 | 22.8 | 12.2 | 12.9 |
| RDPK | 45.2 | 45.9 | 2.6 | 2.7 | 4.0 | 5.0 | 35.1 | 49.3 | 21.7 | 25.7 | 14.8 | 16.1 |
| ERVY | - | - | - | - | - | - | - | - | - | - | - | - |
| FRLL | - | - | - | - | - | - | - | - | - | - | - | - |
| GLND | - | - | - | - | - | - | - | - | - | - | - | - |
| MNKT | 6.9 | 9.1 | 19.9 | 33.2 | 6.3 | 8.8 | 7.2 | 12.3 | 10.1 | 15.8 | -54.4 | -62.2 |
| OPCH | 6.0 | 8.3 | 16.5 | 29.3 | 8.2 | 11.1 | 6.6 | 8.9 | 9.3 | 14.4 | -66.3 | -72.8 |
| A Sand | 4.6 | 6.2 | 16.3 | 26.7 | 5.6 | 10.3 | 5.1 | 6.5 | 7.9 | 12.4 | -83.4 | -89.6 |
| B Dolo | 10.5 | 14.5 | 12.5 | 22.5 | 15.5 | 23.1 | 18.2 | 25.8 | 14.2 | 21.5 | -80.2 | -82.5 |
| B Sand | 10.1 | 14.2 | 12.6 | 22.3 | 15.4 | 22.9 | 18.4 | 25.8 | 14.1 | 21.3 | -86.8 | -88.6 |
| C1 Dolo | 37.7 | 67.6 | 8.1 | 11.2 | 7.7 | 12.1 | 4.3 | 5.1 | 14.5 | 24.0 | -98.8 | -95.7 |
| C1 Sand | - | - | - | - | - | - | - | - | - | - | - | - |
| C2 Dolo | - | - | - | - | - | - | - | - | - | - | - | - |
| C2 Sand | 37.2 | 37.2 | 19.8 | 19.8 | 116.1 | 116.1 | 16.2 | 16.2 | 47.3 | 47.3 | -53.8 | -54.2 |
| C3 Dolo | 23.6 | 23.6 | 12.5 | 12.5 | 23.6 | 23.6 | 2.7 | 2.7 | 15.6 | 15.6 | -94.4 | -94.4 |
| C3 Sand | 38.9 | 38.9 | 18.4 | 18.4 | 137.4 | 137.4 | 20.3 | 20.3 | 53.8 | 53.8 | -53.6 | -54.1 |
| C4 Dolo | 41.3 | 41.3 | 20.6 | 20.6 | 143.8 | 143.8 | 20.5 | 20.5 | 56.5 | 56.5 | -53.8 | -54.3 |
| C4 Sand | 38.0 | 38.0 | 15.4 | 15.4 | 150.8 | 150.8 | 23.6 | 23.6 | 56.9 | 56.9 | -56.7 | -57.1 |
| D Dolo | - | - | - | - | - | - | - | - | - | - | - | - |
| D Sand | - | - | - | - | - | - | - | - | - | - | - | - |
| E Dolo | - | - | - | - | - | - | - | - | - | - | - | - |
| AMSD | - | - | - | - | - | - | - | - | - | - | - | - |
| MDSN | - | - | - | - | - | - | - | - | - | - | - | - |
| PC | - | - | - | - | - | - | - | - | - | - | - | - |

*Positive values mean RecSys outperformed state of the art Green's Function spline. Negative values mean Green's Function spline outperformed RecSys. Value is magnitude of outperformance in meters

## Mannville Group

### Recommender System

| Top Name | Fold 1 Number of Training Samples | Fold 1 Number of Holdout Samples | Fold 1 MAE (m) | Fold 1 RMSE (m) | Fold2 Number of Training Samples | Fold2 Number of Holdout Samples | Fold2 MAE (m) | Fold2 RMSE (m) | Fold 3 Number of Training Samples | Fold 3 Number of Holdout Samples | Fold 3 MAE (m) | Fold 3 RMSE (m) | Fold 4 Number of Training Samples | Fold 4 Number of Holdout Samples | Fold 4 MAE (m) | Fold 4 RMSE (m) | Average MAE (m) | Average RMSE (m) |
|---|---|---|---|---|---|---|---|---|---|---|---|---|---|---|---|---|---|---|
| Mannville Group | 1195 | 396 | 40.0 | 50.7 | 1187 | 404 | 40.1 | 51.1 | 1187 | 404 | 47.1 | 57.4 | 1204 | 387 | 44.2 | 132.6 | 42.9 | 73.0 |
| Transgression 8 | 254 | 74 | 31.0 | 44.2 | 247 | 81 | 40.3 | 56.2 | 246 | 82 | 6.7 | 16.3 | 237 | 91 | 5.8 | 8.3 | 21.0 | 31.3 |
| Transgression 7 | 272 | 90 | 8.6 | 17.6 | 272 | 90 | 11.2 | 23.8 | 276 | 86 | 8.2 | 23.0 | 266 | 96 | 7.2 | 15.1 | 8.8 | 19.9 |
| Transgression 6 | 436 | 152 | 12.9 | 19.1 | 433 | 155 | 11.9 | 25.1 | 460 | 128 | 14.5 | 28.9 | 435 | 153 | 13.9 | 51.8 | 13.3 | 31.2 |
| Transgression 5 | 1422 | 486 | 6.0 | 19.9 | 1436 | 472 | 6.4 | 11.9 | 1432 | 476 | 6.0 | 7.1 | 1434 | 474 | 6.3 | 19.4 | 6.2 | 14.6 |
| Clearwater/Wabiskaw | 319 | 106 | 3.2 | 4.7 | 306 | 119 | 3.6 | 5.4 | 327 | 98 | 4.1 | 8.9 | 323 | 102 | 4.2 | 5.9 | 3.8 | 6.2 |
| Transgression 4 | 1177 | 399 | 3.5 | 22.0 | 1196 | 380 | 3.9 | 16.3 | 1182 | 394 | 3.8 | 13.2 | 1173 | 403 | 5.7 | 37.2 | 4.2 | 22.2 |
| Regression 3 | - | - | - | - | - | - | - | - | - | - | - | - | - | - | - | - | - | - |
| Transgression 3 | 1328 | 455 | 3.3 | 20.9 | 1331 | 452 | 3.9 | 16.3 | 1350 | 433 | 3.6 | 14.0 | 1340 | 443 | 5.3 | 36.7 | 4.0 | 22.0 |
| Regression 2 | 1328 | 455 | 3.6 | 21.1 | 1331 | 452 | 3.9 | 16.4 | 1350 | 433 | 3.5 | 14.1 | 1340 | 443 | 5.1 | 36.7 | 4.0 | 22.1 |
| Transgression 2 | 1390 | 441 | 3.5 | 21.5 | 1379 | 452 | 3.1 | 10.6 | 1372 | 459 | 3.6 | 14.3 | 1352 | 479 | 5.2 | 36.0 | 3.8 | 20.6 |
| Transgression 1 | 1354 | 443 | 4.3 | 21.8 | 1348 | 449 | 4.8 | 15.7 | 1340 | 457 | 4.8 | 14.9 | 1349 | 448 | 6.4 | 37.8 | 5.0 | 22.6 |
| Regression 1 | 1341 | 442 | 4.4 | 21.9 | 1343 | 440 | 4.9 | 15.9 | 1325 | 458 | 4.7 | 14.8 | 1340 | 443 | 6.2 | 37.6 | 5.1 | 22.6 |
| McMurray Formation | 1413 | 467 | 4.8 | 21.6 | 1416 | 464 | 6.8 | 19.4 | 1398 | 482 | 5.8 | 16.2 | 1413 | 467 | 8.2 | 38.5 | 6.4 | 23.9 |
| Paleozoic | 1580 | 519 | 30.2 | 42.8 | 1579 | 520 | 40.6 | 53.3 | 1572 | 527 | 38.5 | 48.5 | 1566 | 533 | 32.5 | 72.8 | 35.4 | 54.4 |

### Green's Function Interpolation

| Top Name | Fold 1 MAE (m) | Fold 1 RMSE (m) | Fold 2 MAE (m) | Fold 2 RMSE (m) | Fold 3 MAE (m) | Fold 3 RMSE (m) | Fold 4 MAE (m) | Fold 4 RMSE (m) | Average MAE (m) | Average RMSE (m) | Difference: Method MAE Difference* | Difference: Method RMSE Difference* |
|---|---|---|---|---|---|---|---|---|---|---|---|---|
| Mannville Group | 20.2 | 63.9 | 21.1 | 95.5 | 102.5 | 522.8 | 26.1 | 308.0 | 42.5 | 247.5 | -0.4 | 174.6 |
| Transgression 8 | 14.4 | 51.8 | 11.8 | 27.7 | 37.0 | 87.4 | 9.1 | 20.5 | 18.1 | 46.9 | -2.9 | 15.6 |
| Transgression 7 | 20.4 | 50.2 | 18.8 | 55.8 | 13.8 | 32.2 | 16.8 | 50.6 | 17.5 | 47.2 | 8.7 | 27.3 |
| Transgression 6 | 11.6 | 49.7 | 7.3 | 13.1 | 7.1 | 9.5 | 7.9 | 16.9 | 8.5 | 22.3 | -4.8 | -8.9 |
| Transgression 5 | 32.8 | 213.2 | 24.8 | 282.9 | 39.1 | 284.4 | 22.0 | 203.6 | 29.7 | 246.0 | 23.5 | 231.5 |
| Clearwater/Wabiskaw | 4.5 | 6.6 | 4.9 | 6.7 | 5.5 | 7.8 | 7.0 | 13.6 | 5.5 | 8.6 | 1.7 | 2.4 |
| Transgression 4 | 21.2 | 62.2 | 40.4 | 254.2 | 50.6 | 288.7 | 28.7 | 312.0 | 35.2 | 229.3 | 31.0 | 207.1 |
| Regression 3 | - | - | - | - | - | - | - | - | - | - | - | - |
| Transgression 3 | 32.4 | 173.5 | 26.7 | 293.4 | 15.0 | 79.9 | 53.1 | 405.7 | 31.8 | 238.1 | 27.8 | 216.1 |
| Regression 2 | 32.5 | 173.5 | 26.8 | 293.4 | 14.9 | 79.7 | 52.9 | 405.5 | 31.8 | 238.0 | 27.8 | 215.9 |
| Transgression 2 | 23.7 | 166.4 | 31.2 | 292.7 | 65.7 | 364.3 | 26.6 | 210.6 | 36.8 | 258.5 | 33.0 | 237.9 |
| Transgression 1 | 11.9 | 36.1 | 27.3 | 293.2 | 24.9 | 212.7 | 36.9 | 261.3 | 25.3 | 200.8 | 20.2 | 178.3 |
| Regression 1 | 12.7 | 38.0 | 29.0 | 297.3 | 23.8 | 210.3 | 37.4 | 261.8 | 25.7 | 201.9 | 20.7 | 179.3 |
| McMurray Formation | 14.2 | 52.1 | 26.1 | 287.0 | 25.8 | 205.7 | 33.6 | 253.9 | 24.9 | 199.7 | 18.5 | 175.8 |
| Paleozoic | 14.5 | 37.6 | 45.1 | 251.8 | 39.6 | 233.6 | 30.0 | 270.3 | 32.3 | 198.3 | -3.1 | 144.0 |

*Positive values mean RecSys outperformed state of the art Green's Function spline. Negative values mean Green's Function spline outperformed RecSys. Value is magnitude of outperformance in meters